\documentclass[iop,numberedappendix]{emulateapj}

\usepackage{url}
\usepackage{natbib}
\usepackage{array}
\usepackage{rotating}
\usepackage{subfigure}

\newcommand{\beq}{\begin{equation}}
\newcommand{\eeq}{\end{equation}}
\newcommand{\bdi}{\begin{displaymath}}
\newcommand{\edi}{\end{displaymath}}
\newcommand{\IRAS}{\textit{IRAS}}

\newcommand{\spitzer}{\textit{Spitzer}}
\newcommand{\herschel}{\emph{Herschel}}

\newcommand{\degree}{$^{\circ}$}

\begin{document}

\shorttitle{\textit{Herschel} Observations of High-Mass Star Formation in the W3 GMC}
\shortauthors{Rivera-Ingraham, A.~et al.}

\title{ \textit{Herschel}\footnote{\MakeLowercase{\textit{\MakeUppercase{h}erschel} is an
      \MakeUppercase{ESA} space observatory with science instruments
      provided by \MakeUppercase{E}uropean-led
      \MakeUppercase{P}rincipal \MakeUppercase{I}nvestigator consortia
      and with important participation from \MakeUppercase{NASA}.}}
  Observations of the W3 GMC: Clues to the Formation of Clusters of
  High-Mass Stars }


\author{A.~Rivera-Ingraham,\altaffilmark{1,2,3}
        P.~G.~Martin,\altaffilmark{4}
        D.~Polychroni,\altaffilmark{5}
        F.~Motte,\altaffilmark{6}
        N.~Schneider,\altaffilmark{6,7,8}
        S.~Bontemps,\altaffilmark{7,8}
        M.~Hennemann,\altaffilmark{6}
        A.~Men'shchikov,\altaffilmark{6}
        Q.~Nguyen Luong,\altaffilmark{4}
        Ph.~Andr\'{e},\altaffilmark{6}
        D.~Arzoumanian,\altaffilmark{6}
        J.-Ph.~Bernard,\altaffilmark{3}
        J.~Di~Francesco,\altaffilmark{9,10}
        D.~Elia,\altaffilmark{11}
        C.~Fallscheer,\altaffilmark{10,9}
        T.~Hill,\altaffilmark{6}
        J.~Z.~Li,\altaffilmark{12}
        V.~Minier,\altaffilmark{6}
        S.~Pezzuto,\altaffilmark{11}
        A.~Roy,\altaffilmark{4}
        K.~L.~J.~Rygl,\altaffilmark{11}
        S.~I.~Sadavoy,\altaffilmark{10,9}
        L.~Spinoglio,\altaffilmark{11}
        G.~J.~White,\altaffilmark{13,14}
        C.~D.~Wilson\altaffilmark{15}}
\altaffiltext{1}{Department of Astronomy and Astrophysics, University of
  Toronto, 50 St. George Street, Toronto, ON M5S~3H4, Canada}
\altaffiltext{2}{Universit\'{e} de Toulouse, UPS-OMP, IRAP, F-31028 Toulouse cedex 4, France}
\altaffiltext{3}{CNRS, IRAP, 9 Av. colonel Roche, BP 44346, F-31028 Toulouse cedex 4, France}
\altaffiltext{4}{Canadian Institute for Theoretical Astrophysics,
  University of Toronto, 60 St. George Street, Toronto, ON M5S~3H8,
  Canada} 
\altaffiltext{5}{Department of Astrophysics, Astronomy and Mechanics, Faculty of Physics, University of Athens, Panepistimiopolis, 15784 Zografos, Athens, Greece}
\altaffiltext{6}{Laboratoire AIM, CEA/DSM/Irfu Ð- CNRS/INSU Ð- Universit\'e Paris Diderot, CEA-Saclay, F-91191 Gif-sur-Yvette Cedex, France}
\altaffiltext{7}{Univ. Bordeaux, LAB, UMR 5804, F-33270 Floirac, France}
\altaffiltext{8}{CNRS, LAB, UMR 5804, F-33270 Floirac, France}
\altaffiltext{9}{National Research Council Canada, Herzberg Institute of Astrophysics, 5071 West Saanich Road, Victoria, BC, V9E 2E7, Canada}
\altaffiltext{10}{Department of Physics and Astronomy, University of Victoria, PO Box 355, STN CSC, Victoria, BC, V8W 3P6, Canada}
\altaffiltext{11}{INAF-Istituto di Astrofisica e Planetologia Spaziali, via Fosso del Cavaliere 100, I-00133 Rome, Italy}
\altaffiltext{12}{National Astronomical Observatories, Chinese Academy of Sciences, Beijing, China}
\altaffiltext{13}{Department of Physical sciences, The Open
University, Milton Keynes, UK}
\altaffiltext{14}{RALspace, The Rutherford Appleton Laboratory, Chilton, Didcot, UK}
\altaffiltext{15}{Department of Physics and Astronomy, McMaster University, Hamilton, ON, L8S 4M1, Canada}

\begin{abstract}

The W3 GMC is a prime target for the study of the early stages of high-mass star formation. We have used Herschel data from
the HOBYS key program to produce and analyze column density and
temperature maps.  Two preliminary catalogs were produced by
extracting sources from the column density map and from \herschel\
maps convolved to the 500\,\micron\ resolution. \herschel\ reveals
that among the compact sources (FWHM$<$0.45\,pc), W3 East, W3 West, and W3 (OH) are the most massive and luminous and have
the highest column density. 
Considering the unique properties of W3 East and W3 West, the only clumps with on-going high-mass star formation, we suggest a `convergent constructive
feedback' scenario to account for the formation of a cluster with decreasing age and increasing system/source mass toward the
innermost regions.
This process, which relies on feedback by high-mass stars
to ensure the availability of material during cluster formation, could also lead to the creation of an
environment suitable for the formation of Trapezium-like systems.
In common with other scenarios proposed in other HOBYS studies, our results indicate that an active/dynamic
process aiding in the accumulation, compression, and confinement of
material is a critical feature of the high-mass star/cluster
formation, distinguishing it from classical low-mass star formation.
The environmental conditions and availability of triggers
determine the form in which this process occurs, implying that
high-mass star/cluster formation could arise from a range
of scenarios: from large scale convergence of turbulent flows, to
convergent constructive feedback or mergers of filaments.

\end{abstract}



\section{Introduction} \label{sec:intro}

W3 is a $\sim4\times10^5$\,M$_\odot$ (\citealp{moore2007};
\citealp{polychroni2012}) Giant Molecular Cloud (GMC) well known for
its rich population of \ion{H}{2} regions, clusters, and high-mass
star forming sites (see e.g., \citealp{megeath2008} for a detailed
description of the field and a review of recent literature).  The
relatively close distance of W3 ($\sim2$\,kpc; e.g.,
\citealp{hachisuka2004}; \citealp{xu2006}; \citealp{navarete2011}) has
made this cloud a prime target for the study of cluster/high-mass star
formation, which compared to the low-mass case is much less well
understood.  The question of whether high-mass star formation is
simply a scaled-up version of low-mass star formation, or if it is the
result of a completely different process (sometimes defined as
`bimodality' in star formation), remains one of the main outstanding
issues in star formation theory (e.g., \citealp{zinnecker2007}).
Furthermore, while low-mass stars are able to form in isolation, most
star formation occurs in clusters embedded in their parent GMCs
\citep{lada2003}.  This situation is particularly true for high-mass
stars, making cluster studies crucial to
investigating and understanding the origin of high-mass stars.

The W3 GMC was observed with \herschel\ \citep{pilbratt2010} as part
of the Guaranteed Time Key Program
HOBYS\footnote{http://www.herschel.fr/cea/hobys/en/} (\herschel\
imaging survey of OB Young Stellar objects; \citealp{motte2010}).  The
program is specifically designed to address the major outstanding
issues in high-mass star formation with the analysis of all major
regions with high-mass stars at distances less than d$\sim3$\,kpc.

The paper presents a first look at the W3 GMC with the recently
acquired HOBYS data, focusing on those structures currently hosting
the youngest high-mass stars in this field.  By identifying and
characterizing the properties and origin of these systems, we aim to
constrain the pre-requisites for the formation of clusters of
high-mass stars.  This study is the first of a series of
\herschel-based papers on W3 currently in preparation, and complements
our previous analysis of the YSO content of this field
\citep{rivera2011} and the study of CO by \citet{polychroni2012} .

This paper is organized as follows.  
An introduction to the W3 GMC is given in Section~\ref{w3gmc}.
Section \ref{sec:data_paper1} introduces the
\herschel\ datasets and images.  A global overview of the properties of W3 and its fields as seen by \herschel\ based on the
column density and temperature maps is
described in Section~\ref{sec:global}.
In Section~\ref{sec:msf}, we focus on the unique properties
characteristic of the high-mass star formation in W3 Main.
Section~\ref{sec:msf-conclusion} outlines a new high-mass formation
scenario, described as 'convergent constructive feedback.'
Our conclusions are summarized in Section~\ref{requisites} .

\subsection{The W3 GMC}\label{w3gmc}

W3 contains high-mass stars in various evolutionary stages (e.g., see
\citealp{tieftrunk1997}). The most active star forming sites are W3
Main, W3 (OH), and AFGL 333. All of these regions (see annotations on
Figure~\ref{fig:nh2-ism} below) are located in a prominent and dense
structure defined in the literature as the `high-density layer' (HDL;
e.g., \citealp{oey2005}).  The HDL forms the western edge of the W4
`Heart Nebula' bubble, powered by eleven O stars in the cluster
IC~1805 (2$^{\mathrm{h}}$\,32$^{\mathrm{m}}$\,42$^{\mathrm{s}}$
  +61$^{\circ}$\,27\arcmin\ (J2000)).

W3 Main contains the most prominent high-mass population of the
entire GMC.  The two brightest and most central infrared sources in
this region, IRS5 and IRS4 \citep{wynn1972}, are the youngest
high-mass systems in the GMC, as indicated by the presence of numerous
hypercompact \ion{H}{2} (HC\ion{H}{2}) regions
\citep{tieftrunk1997}.  IRS5 has been suggested to contain a
proto-Trapezium system (\citealp{megeath2005}; \citealp{rodon2008})
powered by a cluster of OB stars (e.g., \citealp{claussen1994};
\citealp{vandertak2005}).

The W3 (OH) region is also comprised of two main regions: W3 (OH)
itself, a young ultracompact \ion{H}{2} (UC\ion{H}{2}) region rich in
OH masers \citep{dreher1981}, and a hot core with a younger massive
proto-binary system, $\sim6$\arcsec\ east of W3 (OH)
\citep{chen2006}.  The region associated with the binary has prominent
H$_2$O emission and it is commonly referred in the literature as W3
(H$_2$O) \citep{little1977, turner1984}.

W3 (OH) and W3 Main are both located in the shell surrounding the
cluster IC 1795. This $3-5$\,Myr old cluster is powered by various OB
stars, the most massive of which, BD $+61$\degree $411$, has a
spectral type O6.5V \citep{oey2005}. The location of W3 Main and W3
(OH) at the edges of the shell around this cluster has been suggested
as evidence for their formation having been induced by IC 1795, itself
created by an earlier burst of star formation in the W4 region
\citep{oey2005}.  While a triggered origin by IC 1795 appears to be
the case for W3 (OH), the characteristics of the cluster of low mass
stars associated with W3 Main suggest that this region might actually
have begun to form at an earlier stage, prior to IC 1795
(\citealp{feigelson2008}; \citealp{rivera2011}).

KR 140 is a small \ion{H}{2} region, $5.7$\,pc in diameter
\citep{kerton2001}.  Located west of the main star-forming activity in
the W3 complex and $\sim40$\,pc SW of IRS5, this \ion{H}{2} region is
powered by a central and isolated O8.5 star (VES 735) about $1-2$\,Myr
old \citep{ballantyne2000}.  Contrary to W3 Main, KR 140 has been
suggested to be the result of a rare case of spontaneous high-mass
star formation \citep{ballantyne2000}.

North of KR 140 there is a filamentary-like structure called `KR
140-N' \citep{rivera2011} or the `Trilobite'
(\citealp{polychroni2010}; \citealp{polychroni2012}).  Its morphology
and associated population of young stellar objects (YSOs) suggest the
Trilobite could be a case of `Radiative Driven Implosion' (RDI;
\citealp{rivera2011}), in which an ionization/shock front, such as that driven by an expanding \ion{H}{2} region, causes a neighboring overdensity to collapse, triggering the formation of stars.

Several of these prominent features are labeled below in
Figures~\ref{fig:nh2-ism} and \ref{fig:temp-ism}.

\subsection{Constraining the High-Mass Star Formation Process}

Emerging evidence presented in recent studies of several of the HOBYS
fields suggests the importance of dynamical processes in
high-mass/cluster formation, associated preferentially with filamentary-like regions
of high column density of the order of
$N_{\mathrm{H}_2}\sim10^{23}$\,cm$^{-2}$ (`ridges';
\citealp{hill2011}; \citealp{nguyen2011}; \citealp{hennemann2012}). These structures could be formed by convergence of flows,
as proposed for W43 \citep{nguyen2011b} and the DR 21 Ridge in
Cygnus-X, the latter being one of the most prominent examples
(\citealp{schneider2010b}; \citealp{hennemann2012}).  The effects of this
process in turbulent environments have been investigated extensively
in previous studies (e.g., \citealp{maclow2004};
\citealp{klessen2005}; \citealp{heitsch2006}).  Other studies of
high-mass star forming regions suggest that enhanced accretion at the
mergers of filaments is necessary in order to form clusters on short timescales (\citealp{dale2011};
\citealp{schneider2012}). Such a rapid formation process is required to agree with estimates of the lifetime of the prestellar phase of massive YSOs, which may be as short as a single free-fall time (e.g., \citealp{motte2007}).  

Questions remain as to whether such models can successfully explain
the origin of other known high-mass systems in other regions.  If more
than one scenario is needed to explain these systems, it is essential to identify the
conditions common to all that might constitute the basic requisites
for the formation of OB stars and their associated clusters.  We return
to this in Sections~\ref{sec:msf-conclusion} and \ref{requisites}.


\section{Data Processing and \herschel\ Images} \label{sec:data_paper1}

\begin{figure}[ht]
\centering
\includegraphics[scale=0.35,trim=2cm 2cm 0cm 2cm,clip=true]{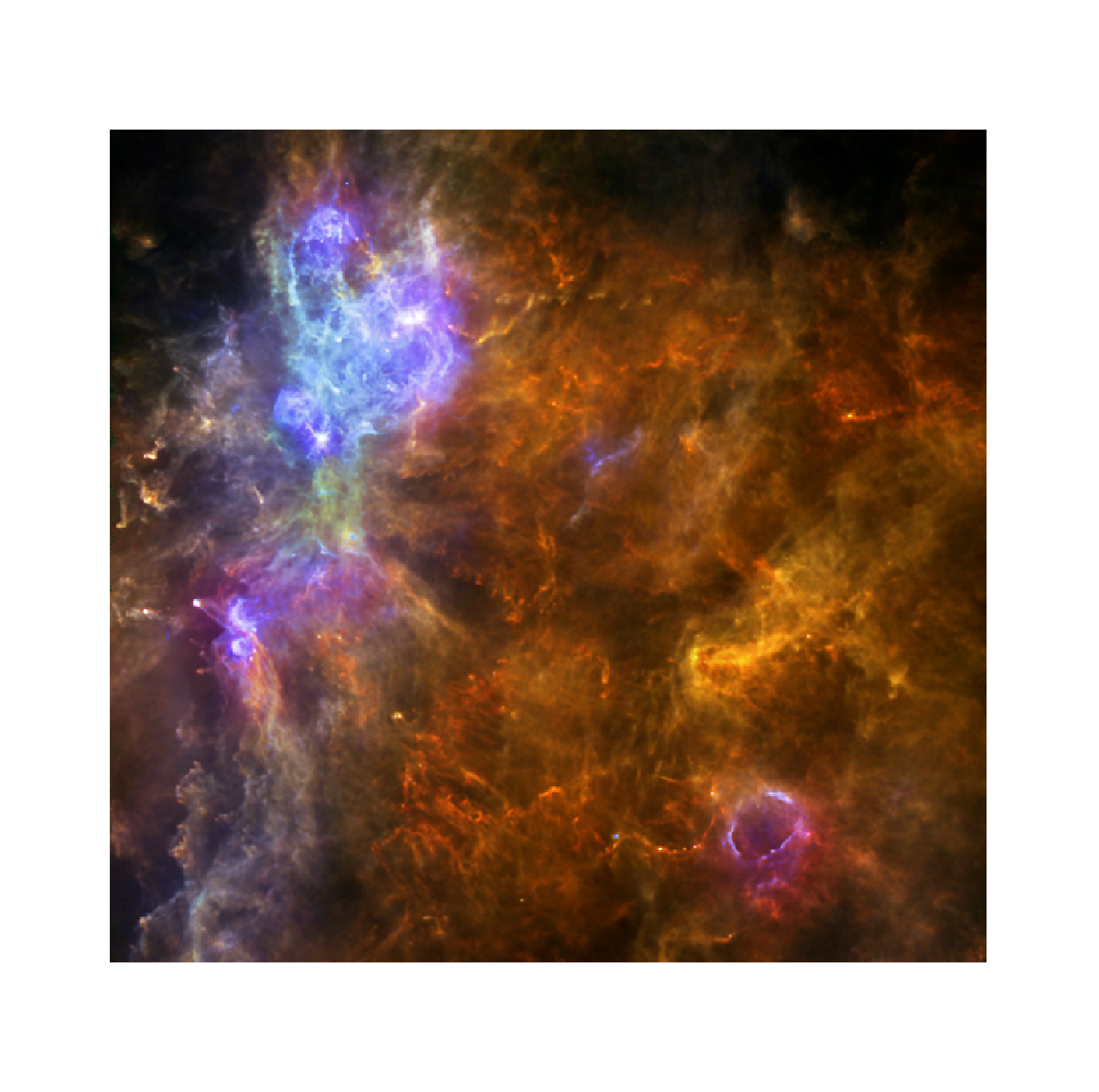}
\caption{Three color image of the W3 GMC using
  \herschel\ HOBYS data at $70$\,\micron (blue), $160$\,\micron
  (green) and $250$\,\micron (red). The color tables have been manipulated to bring out the structural detail in the map.}
\label{fig:herschel}
\end{figure}

W3 was observed by \herschel\ in March 2011.  Continuum data were
obtained in parallel mode at $70$\,\micron\ and $160$\,\micron\ with
PACS \citep{pog2010}, and $250$\,\micron, $350$\,\micron, and
$500$\,\micron\ with SPIRE \citep{griffin2010}.  The most prominent
thermal emission features in W3 are shown in unprecedented detail by
the \herschel\ data.

Datasets were reduced from level 0 to level 1 using HIPE v8.1, and the final maps were
produced with v17.0 of the Scanamorphos\footnote{http://www2.iap.fr/users/roussel/herschel/index.html}
software package \citep{roussel2012}.  Astrometry was corrected using the Two Micron All
Sky Survey\footnote{The Two Micron All Sky Survey (2MASS) is a joint
  project of the University of Massachusetts and the Infrared
  Processing and Analysis Center/California Institute of Technology,
  funded by the National Aeronautics and Space Administration and the
  National Science Foundation.} and \spitzer\
data\footnote{http://irsa.ipac.caltech.edu/applications/Gator/}.
The pointing accuracy is estimated to be better than $2\farcs0$.  Maps
were transformed to MJy/sr and offsets were determined using
\emph{Planck} and \IRAS\ data \citep{bernard2010}.

A small saturated area was detected in W3 (OH) at $70$, $250$, and
$350$\,\micron\
($\sim$2$^{\mathrm{h}}$\,27$^{\mathrm{m}}$\,4$^{\mathrm{s}}$
$+61^{\circ}$\,52\arcmin\,24\arcsec).  This corresponds to the source
IRS8 \citep{wynn1972}.
Other small (a few pixels) regions were also saturated in W3 Main at
$70$\,\micron\
(2$^{\mathrm{h}}$\,25$^{\mathrm{m}}$\,40$^{\mathrm{s}}.5$
$+62^{\circ}$\,05\arcmin\,50\arcsec) and $250$\,\micron\
(2$^{\mathrm{h}}$\,25$^{\mathrm{m}}$\,40$^{\mathrm{s}}.5$
$+62^{\circ}$\,05\arcmin\,50\arcsec;
2$^{\mathrm{h}}$\,25$^{\mathrm{m}}$\,30$^{\mathrm{s}}.5$
$+62^{\circ}$\,06\arcmin).  These correspond to the sources IRS5 and
IRS4, respectively \citep{wynn1972}.  
These regions were reobserved with \herschel\ using the `bright source' mode.
These images showed a good correlation with the originals in the
unsaturated areas of overlap and were used to replace the intensities
in the saturated pixels.

A three-color image of W3 is shown in Figure \ref{fig:herschel}. 
The five individual continuum images are presented in
Appendix~\ref{5images}.
Our analysis focuses on the `common survey area',
$\sim$1.5\,$^{\circ}$ in extent in RA and Dec,
scanned in two roughly orthogonal directions by PACS and SPIRE
according to the telescope focal plane orientation on the date of
observation. 

Supplementary Stokes I continuum images at 1420 MHz from the Canadian
Galactic Plane Survey (CGPS; \citealp{taylor2003}) were used to
investigate the location and distribution of the ionized regions.


\section{Column Density and Dust Temperature Maps}\label{sec:global}

\subsection{Data Analysis}\label{sec:maps}

\begin{figure*}[ht]
\centering
\includegraphics[scale=0.53,angle=270]{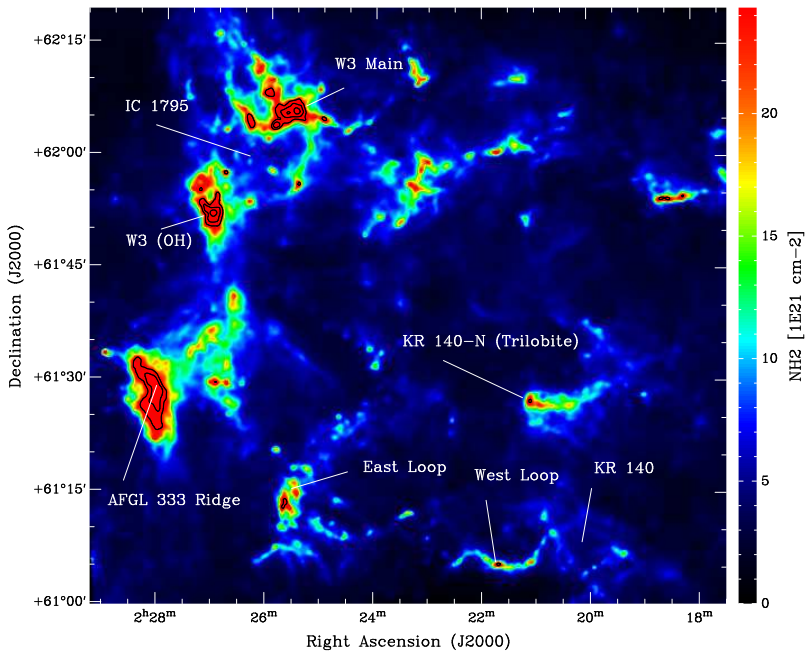}
\caption{Column density map of the W3 GMC after
  correction for dust emission associated with foreground/background
  atomic and molecular material, as described in
  Appendix~\ref{foreground}. A variety of filaments, pillars and
  structures are found throughout the GMC. Labels mark prominent
  features in W3 (Section~\ref{w3gmc}). The `HDL' is the dense region comprising W3 Main, W3 (OH), and the AFGL 333 Ridge. Contours are $N_{\mathrm{H}_2}
  \approx [30, 60, 200] \times10^{21}$\,cm$^{-2}$.}
\label{fig:nh2-ism}
\end{figure*}

\begin{figure*}[ht]
\centering
\includegraphics[scale=0.53,angle=270]{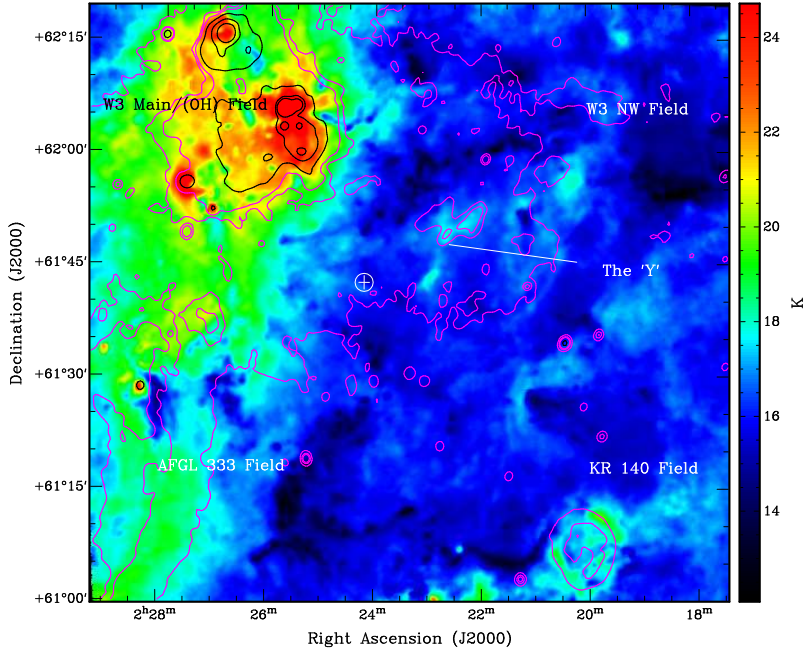}
\caption{Dust temperature map of the W3 GMC after correction
for dust emission associated with foreground/background atomic and
molecular material, as described in Appendix~\ref{foreground}.
Contours are of Stokes I continuum at 1420 MHz: T$_{\mathrm{b}}=8, 12,
15$\,K (magenta), and T$_{\mathrm{b}}=30, 100,$ and $240$\,K
(black). Colors for contours have been chosen for better contrast in
cold and warm regions.
White circle with cross marks the intersection of the four fields in
W3. From left to right and top to bottom: W3 Main/(OH) field, W3-NW,
AFGL 333, KR 140 fields. Figure includes labels for the four fields
and the location of the `Y'-shaped hot structure (see text).}
\label{fig:temp-ism}
\end{figure*}

To derive column density and dust temperature maps, the PACS and
SPIRE maps were first convolved to the resolution at 500\,\micron\,
($\sim$36\arcsec; $\sim$0.35\,pc at a distance of 2\,kpc) and
projected onto a common 9\,\arcsec\,pix$^{-1}$ grid (that of the $500$\,\micron\ map).

The intensities at each pixel were fitted with the
approximation\footnote{At the column density peaks
  (Section~\ref{sec:msf}), the optical depth at 160\,\micron\ exceeds
  0.3.
It is of course even higher at 70\,\micron, but those data are not
used in our assessment.}
\begin{equation}
I_\nu = (1 - \exp(-\tau_\nu))\, B_\nu(T),
\label{column}
\end{equation}
where $B_{\nu}(T)$ is the Planck function for dust temperature $T$.
The common assumption in such studies is that $T$ is constant along
the line of sight, obviously a great simplification given the variety
of structures and radiation fields.  Thus only \herschel\ bands $\ge
160$\,\micron\ were considered in creating the maps.  The optical
depth is parametrized as $\tau_\nu =\tau_{\nu_0} (\nu/\nu_0)^{\beta}$
with $\nu_0 = 1200$~GHz ($250$\,\micron).  The optical depth is
related to the total H column density by $\tau_\nu = \sigma_\nu
N_{\mathrm{H}}$ through the opacity $\sigma_\nu$.

We assumed a fixed dust emissivity index of $\beta=2$ and adopted the
same opacity standard as in other HOBYS and Herschel-based studies (e.g., \citealp{motte2010}; \citealp{andre2010}), 0.1~cm$^2$~gm$^{-1}$ at 1~THz, equivalently $\sigma_0=3.4 \times 10^{-25}$\,cm$^2$\,H$^{-1}$ using 1.4 as the mean atomic weight per H nucleon. This can be used to deduce the total H column density from $\tau$. However, again to be consistent with the scale of column densities in these earlier papers, we used 2.33 rather than 2.8 as the mean atomic weight per molecule, so that what is called $N_{\mathrm{H}_2}$ is really the column density of `mean molecules' (see also \citealp{kauffmann2008}). There is no scaling effect on the mass surface density or the masses of clumps. More importantly, we note that intrinsic uncertainties in the adopted opacity might alter the derived column densities systematically by a factor of $\sim2$, and that there might be systematic differences in opacity across the field or with column density \citep{martin2012}. Compared to this, photometric calibration errors are relatively small, 15$\%$ and 20$\%$ for SPIRE and PACS $160$\,\micron, respectively (\citealp{griffin2010}; \citealp{pog2010}).  

Our goal is to assess the column densities and temperature within the
W3 GMC itself.  The \herschel\ intensities $I_\nu$, however, contain
contributions from dust in the foreground and background, each with
its own $\sigma_\nu$, $T$, and $N_{\mathrm H}$; the right hand side of
Equation~[\ref{column}] is summed over all components.
While a challenging exercise, educated subtraction of the non-GMC
components is advantageous for providing a more accurate
representation of the true local conditions in the GMC.  The process
of subtraction is described in Appendix~\ref{foreground}.
Subtracting the foreground and background is most important for
characterization of regions where the column density in the GMC is
relatively low; it has little effect on the regions of highest column
density.

The final corrected column density map is shown in Figure
\ref{fig:nh2-ism}. The accompanying
temperature map is shown in Figure~\ref{fig:temp-ism}. Both figures
label the most prominent features of the W3 GMC.  Unless mentioned
otherwise, we used these corrected maps as the default images for our
analysis.

For purposes of comparison and discussion, we separate the GMC into
four different `fields,' labeled according to their physical location
with respect to the center of the entire cloud or a major feature
present in the field.  The fields are: W3 Main/(OH), W3~NW, AFGL 333,
and KR 140 field (see Fig. \ref{fig:temp-ism}).  The W3 Main/(OH) and
AFGL 333 fields together comprise what we define as the `eastern' (or
HDL) fields in W3, while the other two are the `western' (`quiescent'
and more `diffuse') ones.

\subsection{Global Overview}\label{overview}

\begin{figure}[ht]
\centering \includegraphics[scale=0.53,angle=0,trim=0.4cm 0cm 0cm
  0cm,clip=true]{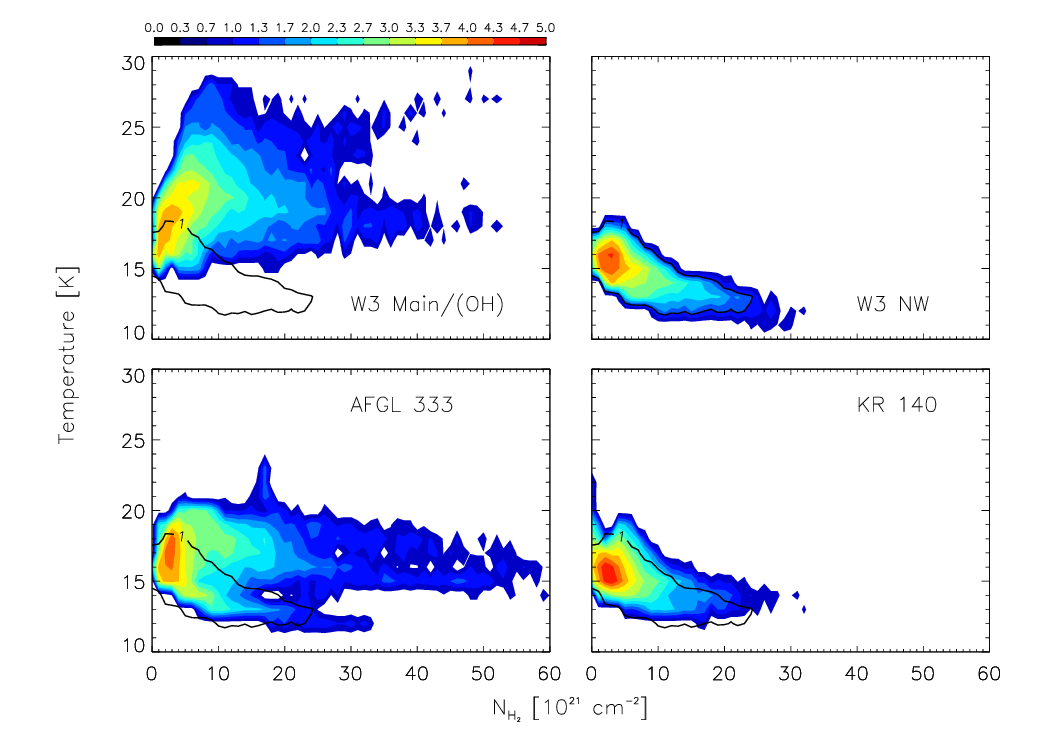}
\caption{Two-dimensional histogram of dust temperature and column
  density in each W3 field. Color bar indicates number of pixels (log
  units). Black contour marks the distribution of the W3~NW field as a
  reference for the distribution in the other fields.}
\label{fig:coltemp_histo_ism2}
\end{figure}

Figure~\ref{fig:coltemp_histo_ism2} presents the relationship between
column density and dust temperature for the four fields, in the form
of a two-dimensional histogram.  The W3 Main/(OH) field ranges to
higher temperatures not observed in the cooler and more diffuse
western fields (W3~NW and KR 140).  The AFGL 333 field shows a similar
high-temperature extension, associated with material at the boundary
with W4, but with peak temperatures between those in W3 Main/(OH) and the
western fields.  In this field there is also low-temperature material,
not present in the W3 Main/(OH) field, associated with the structures of the
East Loop (Fig. ~\ref{fig:nh2-ism}), and with properties consistent with those observed in the W3~NW and KR 140 fields.

While most regions in W3 show a clear anticorrelation between high column densities and low temperatures, the Main/(OH) field is unique due to the association of high column densities with warmer temperatures. These structures, warmed by embedded OB stars, are also associated with the youngest high-mass activity in W3 (e.g., W3 Main). This illustrates the difficulty of using a simple temperature/column density threshold to identify the youngest stages of high-mass star formation in large scale surveys.

Taking into account the material at all temperatures, we find that the
four fields, though of unequal area within the common science region,
have comparable mass.  The main parameters for the four fields, including masses, areas, and surface densities, $\Sigma$, for each of the four fields are shown in Table
\ref{table:global_fields}.
This also includes high column density material above the limits
displayed in Figure~\ref{fig:coltemp_histo_ism2}, which belongs mainly
to the W3 Main/(OH) field.

\begin{table}
\caption{Global Parameters for the W3 Fields}
\label{table:global_fields}
\centering
\begin{tabular}{l l l l l l}
\hline \hline
Field&Mass$_{unc}$$^a$&Mass$^b$&Area&$\Sigma$$^a$&$\Sigma$$^b$\\ 
&[10$^4$\,M$_{\odot}$]&[10$^4$\,M$_{\odot}$]&[deg$^2$]&[M$_{\odot}$/pc$^2$]&[M$_{\odot}$/pc$^2$]\\ 
\hline
Main/(OH)&$7.2$&$6.0$&$0.43$&$136$&$114$\\ 
AFGL 333&$8.1$&$6.3$&$0.50$&$133$&$103$\\ 
KR 140&$9.4$&$5.9$&$0.74$&$104$&$65$\\
NW&$6.5$&$4.8$&$0.53$&$101$&$76$\\ 
\hline
\multicolumn{6}{l}{{$^a$ Mass and surface density from uncorrected (original) maps.}}\\ 
\multicolumn{6}{l}{{$^b$ Mass and surface density corrected for foregr./backgr. material.}}\\ 
\end{tabular}
\end{table}

The total (all fields) corrected mass of the W3 GMC is found to be $\sim2.3\times10^5$\,M$_{\odot}$ ($\sim3.1\times10^5$\,M$_{\odot}$, uncorrected for foreground-background ISM contribution). This estimate is about half that derived by \citet{polychroni2012} from molecular data.  \citet{moore2007} found
$\sim3.8\times10^5$\,M$_{\odot}$ from $^{13}$CO for a constant
excitation temperature of $\sim30$\,K.  Our mass estimates would be
raised (lowered) systematically by adopting a lower (higher) opacity.

\subsection{Stellar Influence and Cloud Structure}\label{central}

The ionizing and eroding activity of the O stars powering W4 is
clearly observed along the eastern edge of our mapped field. The
ionization front can be traced in the radio continuum, as shown with
Stokes I continuum contours in Figure~\ref{fig:temp-ism}.  Diffuse
radio emission is observed to be most prominent in the HDL, with the
strongest peaks coincident with W3 Main, and decreasing progressively
into the colder western fields.

\begin{figure}[ht]
\centering
\includegraphics[scale=0.4,angle=270]{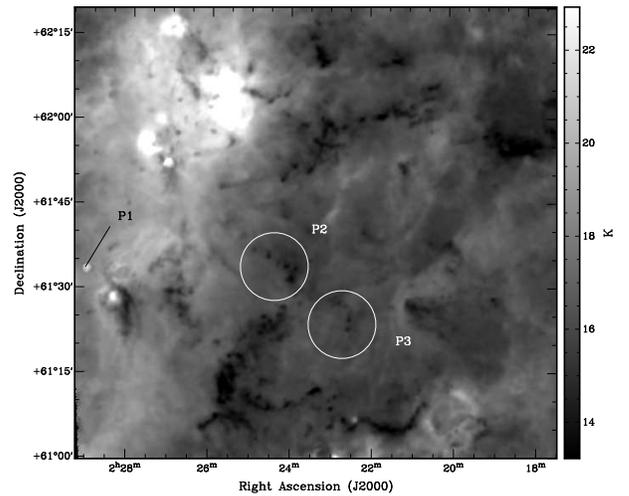}
\caption{Same as Fig. \ref{fig:temp-ism}, but with intensity and color range
  modified to highlight the coldest structures. Circles mark the
  location of pillar-like structures. The IC~1805 cluster is located east of the AFGL 333 Ridge, at 2$^{\mathrm{h}}$\,32$^{\mathrm{m}}$\,42$^{\mathrm{s}}$
  +61$^{\circ}$\,27\arcmin\, (J2000).}
\label{fig:temp-cold}
\end{figure}

Various elongated, dense cloud structures emerging parallel to one another from
the HDL are observed to extend more than 10~pc in projection along the
boundary with W4, pointing toward the east and the O-star cluster
IC~1805 (see Fig. \ref{fig:temp-cold}). The most prominent of these structures (defined as `pillars' in this work), is located just northeast of AFGL 333 and has signatures of ongoing
star formation \citep{rivera2011}.  The location of this structure
(P1) and other pillar-regions in W3 are marked below in
Figure~\ref{fig:temp-cold}, which shows several to be part of the
coldest structures in the W3 GMC. 

There are groups of south-oriented pillars in the central regions of W3, marked P2 and P3 in Figure~\ref{fig:temp-cold} and seen in more detail in Figure~\ref{fig:herschel} and the original single-band Herschel images in Appendix~\ref{5images}.  Interaction with W4 alone cannot account for the level of complexity and diversity of such features found west of the HDL; instead they are suggestive of more local stellar influence in the area. Further morphological details and stellar content of the centrally-located (P2, P3) and
W4-facing pillars were examined with \spitzer\ images by
\citet{rivera2011}.

Contrary to the HDL and eastern fields, stellar influence in the
western fields is very localized, as shown in
Figure~\ref{fig:temp-ism}.  Examples of these `hot' spots are the KR
140 \ion{H}{2} region and a `Y' shape structure
(Fig. \ref{fig:temp-ism}) at $\sim$
2$^{\mathrm{h}}$\,22$^{\mathrm{m}}$\,37$^{\mathrm{s}}$
+61$^{\circ}$\,49\arcmin\,41\arcsec.  
A bow-shock shaped `high' temperature feature wraps around the eastern
side of KR 140-N (Trilobite); displacement in the direction of the
YSOs associated with this structure supports a possible case of
triggered RDI origin \citep{rivera2011}.


\section{Clues to the Elusive High-Mass Star Formation Process: The Case of W3 Main}\label{sec:msf}

With the presence of ongoing high-mass star formation and
$N_{\mathrm{H}_2}$ of the order of $\sim10^{23} $\,cm$^{-2}$, 
the W3 GMC offers a favorable opportunity to characterize rare
high-column density structures and understand how their properties,
origin, and evolution might be linked to the onset of high-mass
activity.

\subsection{Identification of High-Column Density Structures}

\begin{figure}[ht]
\centering
\includegraphics[scale=0.40,angle=270]{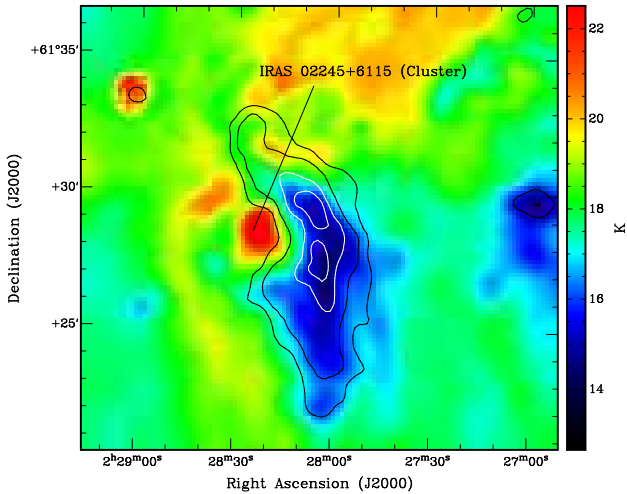}
\caption{Temperature map of the structural details around
  the central regions of the AFGL 333 field.  Black contours are
  $N_{\mathrm{H}_2} \approx [20, 30] \times10^{21}$\,cm$^{-2}$.
  White contours are $N_{\mathrm{H}_2} \approx [60,
  100] \times10^{21}$\,cm$^{-2}$. }
\label{fig:triggered_afgl}
\end{figure}

Figure \ref{fig:triggered_afgl}
shows the temperature map of what we define as the `AFGL 333 Ridge'
(Fig. \ref{fig:nh2-ism}), in the innermost regions of the AFGL 333
field.  It is cold ($<$T$>$$\sim15$\,K, T$_{min}\sim14$\,K, for $N_{\mathrm{H}_2} > 6\times10^{22}$\,cm$^{-2}$). The AFGL 333 Ridge is also the only
filamentary-shaped structure in the GMC reaching peak column densities
of $\sim10^{23}$\,cm$^{-2}$.  This makes it similar to those filaments
defined as `ridges' in previous HOBYS papers (e.g.,
\citealp{hill2011}).

\begin{figure}[ht]
\centering
\includegraphics[scale=0.4,angle=270]{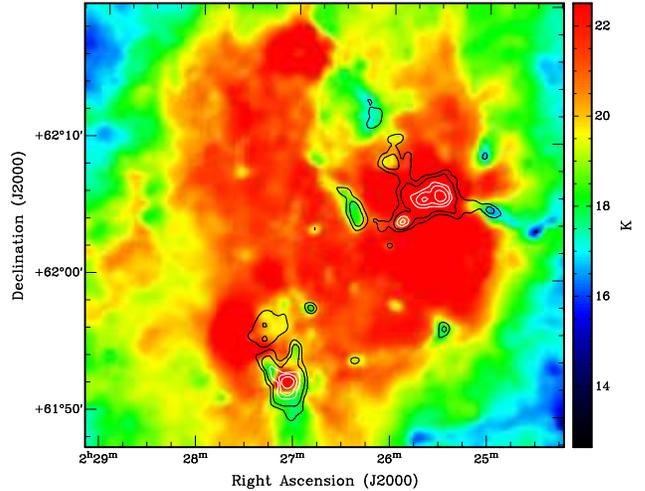}
\caption{Same as Fig. \ref{fig:triggered_afgl}, but for W3 Main, W3 (OH), and W3 North with an additional (white) contour at $N_{\mathrm{H}_2} \approx 200\times10^{21}$\,cm$^{-2}$.}
\label{fig:triggered_main}
\end{figure}

\begin{figure}[ht]
\centering
\includegraphics[scale=0.4,angle=270]{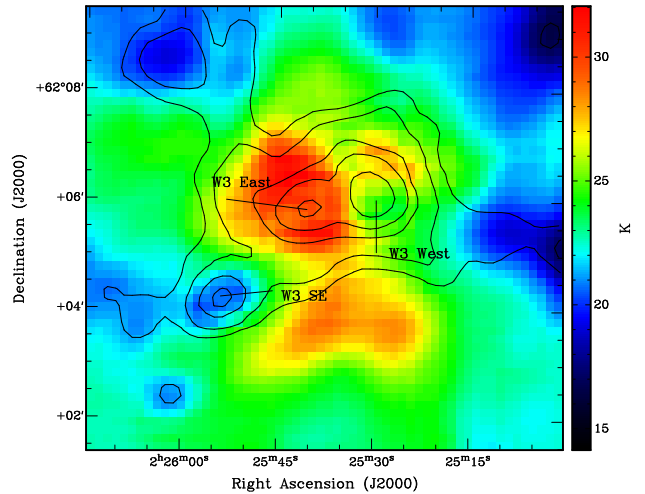}
\caption{Same as Fig. \ref{fig:triggered_main}, but focusing on the W3
Main region.}
\label{fig:w3main}
\end{figure}

\citet{km2008} predicted a star formation threshold for high-mass
stars of $\Sigma=0.7$\,g\,cm$^{-2}$ for a star with $M \sim
10$\,M$_{\odot}$. This surface density corresponds to
$N_{\mathrm{H}_2}\sim1.8\times10^{23}$\,cm$^{-2}$ in our maps.
Although theirs is a specific model, we will for convenience refer to
this column density as a `massive star formation threshold' (MSFT).
Only three structures in the W3 GMC reach column densities of this order, the W3
East and W3 West clumps (FWHM$\sim0.45$\,pc) in W3 Main and W3~(OH) (FWHM$\sim0.43$\,pc).
Compared to the AFGL~333 Ridge, these are all warm ($<$T$>$$\sim25$\,K, T$_{min}\sim18$\,K, for $N_{\mathrm{H}_2} > 6\times10^{22}$\,cm$^{-2}$) as shown in
Figures~\ref{fig:triggered_main} and \ref{fig:w3main} which
concentrate on these most dense regions in the GMC.

The highest column density is in W3
(OH), $N_{\mathrm{H}_2}\sim[4.5\pm1.0] \times10^{23}$\,cm$^{-2}$, comparable within errors to that of W3 West, $N_{\mathrm{H}_2}\sim[3.3\pm0.8] \times10^{23}$\,cm$^{-2}$.

Column densities of this order are consistent with the findings reported by
\citet{white1983}, \citet{richardson1989}, and the presence of a
strong, highly variable water maser in the W3 (OH) region
\citep{little1977}.  W3 East is found to have a peak column density $1.4$ times lower than W3 West ($N_{\mathrm{H}_2}\sim[2.3\pm0.6] \times10^{23}$\,cm$^{-2}$), which agrees with the column density ranking of \citet{tieftrunk1998} based on molecular C$^{18}$O data. We note, however, that these authors predict a column density estimate for W3 (OH) from NH$_{3}$ comparable to that of W3 East.

W3 East, W3 West, and W3 (OH) are the only clumps with confirmed
ongoing (clustered) high-mass star formation in the W3 GMC, as shown
by the presence of masers, HC\ion{H}{2}, or UC\ion{H}{2} regions.  While
the MSFT value is used in this work just as a point of reference when
ranking column densities, if such a threshold holds in practice, then
a total combined mass of $\sim2600$\,M$_{\odot}$ for $N_{\mathrm{H}_2}\ge$\,MSFT, about
$\sim1$\% of the mass of the W3 GMC, is at present possibly
associated with high-mass star formation.

\subsection{Masses and Luminosities}

As a first reconnaissance, we ran the multi-scale, multi-wavelength
source extraction software \textit{getsources}
(\citealp{getsources2012}; v.1.120828) on the column density map (Fig.~\ref{fig:nh2-ism}) made from the \herschel\ images.  Some
properties of the three most massive clumps associated with the
highest column density peaks are summarized in Table \ref{table:main},
assuming a distance of 2.0~kpc.  Because \textit{getsources} estimates
the local background, the corresponding values above the background, as well as the errors as estimated by \textit{getsources}, 
are presented.  The contributions to the luminosity within the
footprint of the structure are evaluated using the mass and $T$
(Fig.~\ref{fig:temp-ism}) associated with each pixel.

\begin{table*}[t!]
\caption{The high extinction structures in the W3 Main/(OH) field: parameters from the $N_{\mathrm{H}_2}$ and $T$ maps}
\label{table:main}
\centering
\begin{tabular}{l l l l l l l}
\hline
\hline
Name&RA&Dec&Peak $N_{\mathrm{H}_2}$$^a$&Peak $T^b$&Mass&$<$FWHM$>$$^c$\\
&[h m s]&[$\circ\,\arcmin\,\arcsec$]&[$10^{23}$\,cm$^{-2}$]&[K]&[10$^3$\,M$_{\odot}$]&[pc]\\
\hline
W3 West&02 25 30.1&62 06 07&$2.9\pm0.8$&$27.2\pm2.2$&$1.54\pm0.03$&0.45\\
W3 East&02 25 40.8&62 05 52&$1.8\pm0.6$&$30.4\pm3.9$&$0.87\pm0.03$&0.42\\
W3 (OH)&02 27 03.7&61 52 21&$4.1\pm1.0$&$25.0\pm1.9$&$1.85\pm0.03$&0.42\\
\hline
\multicolumn{7}{l}{{$^a$ Measured at the coordinate center above the
    local \textit{getsources}-estimated background}}\\
\multicolumn{7}{l}{{$^b$ Measured at the coordinate center.}}\\
\multicolumn{7}{l}{{$^c$ Geometric mean FWHM of elliptical footprint.}}\\
\end{tabular}
\end{table*}

In fitting a modified blackbody, analogous to Equation~\ref{column}, to the
spectral energy distributions (SEDs) from these data, we used $\beta =
2$ and found the temperature $T$, the luminosity $L$, and mass $M$
using the aforementioned opacity. The resulting parameters from our `cold' SED fits ($\lambda\ge160$\,\micron), and their uncertainties as estimated from Monte Carlo simulations, are included in Table
\ref{table:main2}.

\begin{table*}[t!]
\caption{The high extinction structures in the W3 Main/(OH) field: parameters from SED fitting$^a$}
\label{table:main2}
\centering
\begin{tabular}{l l l l l l l l}
\hline
\hline
Name&RA&Dec&Mass&$L$&$T$&$<$FWHM$>^b$&offset$^c$\\
&[h m s]&[$\circ\,\arcmin\,\arcsec$]&[10$^3$\,M$_{\odot}$]&[$10^4$L$_{\odot}$]&[K]&[pc]&[$\arcsec$]\\
\hline
W3 West&02 25 30.5&62 06 13&$1.7\pm0.5$&$3.5\pm1.8$&$24.9\pm3.2$&0.46&6.8\\
W3 East&02 25 40.6&62 05 53&$0.8\pm0.3$&$10.3\pm9.1$&$32.6\pm6.2$&0.42&1.9\\
W3 (OH)&02 27 03.8&61 52 23&$1.6\pm0.5$&$2.3\pm1.0$&$23.3\pm2.7$&0.42&2.0\\
\hline
\multicolumn{8}{l}{{$^a$ Using \textit{getsources} parameters from maps convolved to 500\,\micron\ resolution.}}\\
\multicolumn{8}{l}{{$^b$ Geometric mean FWHM of elliptical
    aperture at $500$\,\micron.}}\\
\multicolumn{8}{l}{{$^c$ Distance from the $N_{\mathrm{H}_2}$ peak
    listed in Table \ref{table:main}}}\\
\end{tabular}
\end{table*}

Ideally one would measure the flux densities by running
\textit{getsources} on the original \herschel\ images. However, our choice to use the 500\,\micron\ resolution for all maps avoids the use of the flux density scaling recipe, used in other HOBYS fields to correct flux
densities for the different resolutions prior to fitting the SED, but whose use is only applicable for cores under certain conditions and physical characteristics (\citealp{motte2010}; \citealp{nguyen2011}).

All of these estimates are in agreement and confirm that the masses of the $\sim0.4$\,pc W3 (OH), W3 West, and W3 East dense clumps are of the order $\sim10^3$\,M$_{\odot}$. These
are also in agreement with previous estimates (e.g.,
\citealp{campbell1995}; \citealp{tieftrunk1998}; \citealp{megeath2008} and references therein).

Regardless of the catalog used (that derived from the column density map or the multiband maps) the \herschel\ data indicate that the clumps currently hosting the on-going high-mass star formation of W3 (W3 East, W3 West, and W3
(OH)) are the most massive and luminous of the entire GMC.  
Only two sources from the column density source catalog, located in the AFGL 333 Ridge, have a
mass of this order (greater than W3 East, albeit less than W3 (OH) and W3 West). They have, however, a lower peak column density and a mean FWHM twice that of the three high extinction sources. Similarly, the only other detection (in addition to W3 (OH) and W3 West) from the convolved multiband catalog with a mass greater than that of W3 East (also associated with the AFGL 333 Ridge), also shows a mean
FWHM$_{\mathrm{500\micron}}\sim1.5$ times larger. There are no reliably-measured sources with a luminosity reaching that of W3 East, W3 West, or W3 (OH).

\subsection{Stellar Content in a \textit{Herschel} Context}

The identification of properties \textit{exclusive} to W3 East, W3
West, and W3 (OH) (structural and stellar content) might reveal clues about their origin and about the general process of high-mass star and cluster
formation.  Therefore, we attempt here to find some such properties by
comparing the \herschel-derived results with characteristics of the extensively studied stellar population of the sources in W3 Main (W3 East and W3 West), the only sources in the W3 GMC with recent/ongoing high-mass star formation (containing HC\ion{H}{2} regions).

\begin{figure}[ht]
\centering
\includegraphics[scale=0.5,angle=0]{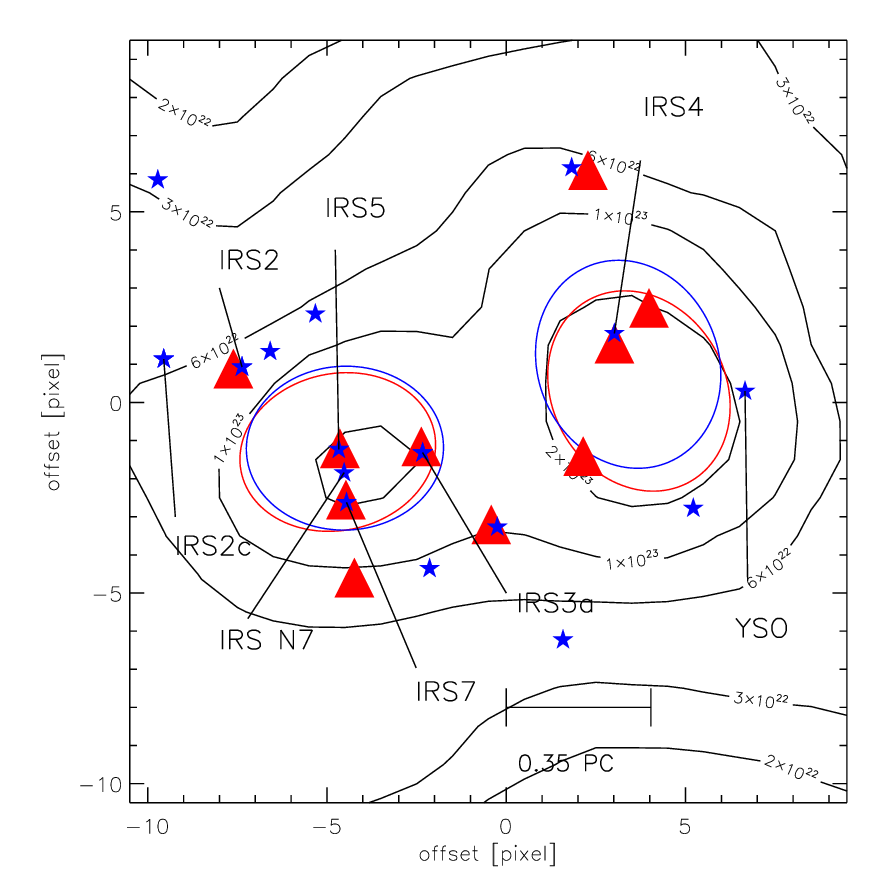}
\caption{Distribution of \ion{H}{2} regions (triangles;
  \citealp{tieftrunk1997}) and OB stars (blue stars;
  \citealp{bik2011}) in W3 Main. Contours are the same as in
  Fig. \ref{fig:w3main}. Ellipses are
  the FHWM ellipses provided by \textit{getsources} during source
  extraction on the convolved \herschel\ maps (blue) and on the column
  density map (red). Axis units are in pixels ($9$\arcsec) offset from
  position
  RA/Dec:2$^{\mathrm{h}}$\,25$^{\mathrm{m}}$\,35$^{\mathrm{s}}$.5
  +62$^{\circ}$\,05\arcmin\,59\arcsec. Labels mark the location of
  high-mass stars mentioned in the text.}
\label{fig:main_fwhm}
\end{figure}

Both clumps are unique in the W3 GMC due to their large population of
young high-mass stars (e.g., \citealp{tieftrunk1997}).  These stars
are present not only within the clumps, but also surrounding their
outer boundaries.  This can be observed in Figure~\ref{fig:main_fwhm},
which shows the location of the OB population with respect to the
\herschel\ detections.  The spectral types of these stars were
analyzed in detail by \citet{bik2011}. 

\subsubsection{W3 East}
W3 East is the most active of the two in terms of the associated
(external and internal) stellar population. 
Using the technique of \citet{rivera2010} we estimate that the
luminosity of the W3 East clump is equivalent to a single-star ZAMS spectral type \citep{panagia1973} of $\sim$O7.
However, because we did not use the $70$\,\micron\ datum in fitting the SED, the temperature and luminosity of this hot source are poorly constrained, and so the allowed range of spectral type is wide, B0.5 to O6. When including the $70$\,\micron\ measurement the SED is much more tightly constrained, with the result T$\approx39$\,K, M$\approx560$\,M$_{\odot}$, L$\approx[1.5\pm0.2] \times10^5$\,L$_{\odot}$, and an implied spectral type of O6.5.
Use of the luminosity Class V table in \citet{martins2005} would make the type `later' by $\sim0.5-1$.
The YSO and proto-Trapezium system IRS5 is located at the column
density peak.  \citet{bik2011} provided no spectral type estimate for
IRS5, but our value is compatible with that of IRS3a, the earliest
reported star within the FWHM of W3 East, with an estimated spectral
type of O5--O7.
Also located within the FWHM (boundary), offset $\sim 0.1$\,pc from the column density peak, is IRS7, a
late O/early B star associated with an UC\ion{H}{2} region and, closer
still to the peak ($\sim0.05$\,pc), IRS N7 (Fig., \ref{fig:main_fwhm}), also a YSO but
older than IRS5 \citep{bik2011}.

Despite the significant population of high-mass stars within its
boundaries, W3 East shows evidence of external heating. The highest
dust temperatures of the entire GMC ($\sim32$\,K) are located
\textit{outside} the NE and SW boundary of the $N_{\mathrm{H}_2}$
clump (Fig. \ref{fig:w3main}), and are coincident with the many late
O/early B stars $\lesssim0.5$\,pc from the peak
(Fig. \ref{fig:main_fwhm}). These outer stars range in spectral type
from early B to O6.5 (IRS2) \citep{bik2011}.

\subsubsection{W3 West}

Despite having more mass and higher column density than W3 East, W3
West has a lower dust temperature and is relatively more quiescent.
High-mass star formation appears not to have yet progressed or been
initiated in the innermost regions; there is no indication of internal
high-mass star phenomena coincident with the \herschel\
N$_{\mathrm{H}_2}$ peak.
Corroborating this, there is an NH$_3$ peak at the position of the
column density peak \citep{tieftrunk1998}, and this peak is offset
from any PACS (hot) infrared source or any \ion{H}{2} region.
By contrast, the column density peak of W3 East lacks significant
NH$_3$ emission suggesting a more advanced state of evolution in which
star formation has already influenced the parental cloud locally
\citep{tieftrunk1998}.

Only IRS4 is found within the boundaries of the W3 West clump,
$\sim0.1$\,pc from the $N_{\mathrm{H}_2}$ peak
(Fig. \ref{fig:main_fwhm}).  It is a B0.5--O8 star predicted to be as
young as IRS5 based on the presence of HC\ion{H}{2} regions
\citep{bik2011}.  This is in excellent agreement with the single-star
main-sequence (ZAMS) spectral type B0.5 (O9.5) derived from the \herschel\
luminosity of $L \sim[3.5\pm1.8] \times10^4$\,L$_{\odot}$.

The other members of the young population of W3 West lie close to or
beyond the boundary of the column density structure
(Fig. \ref{fig:main_fwhm}). The location of a YSO and other high-mass stars toward the southern and SW boundaries could suggest interaction with the diffuse \ion{H}{2} regions at the south.

\subsection{High Column Density Structures Lacking High-Mass Star Formation Indicators}\label{highcolumn}
In addition to W3 (OH), \herschel\ identifies only two other
structures in the W3 GMC with column densities of the order
$N_{\mathrm{H}_2}\sim10^{23}$\,cm$^{-2}$: W3 SE
(Fig. \ref{fig:w3main}), in the W3 Main region, and the AFGL 333
Ridge (Fig. \ref{fig:triggered_afgl}).

W3 SE is the coolest of the three clumps in W3 Main. It is located
$\sim1.3$\,pc from IRS5, with the closest high-mass star indicator being a
diffuse \ion{H}{2} region $<1$\,pc to the southwest.

Also located in the HDL, the AFGL 333 Ridge, despite being at the
boundary with W4, shows evidence of a more locally triggered
origin. For example, it has:

i) an elongated morphology on the east curved around an embedded
cluster \IRAS~02245+6115 (this cluster is $\sim1$\,pc from the
strongest column density peak in the Ridge and contains a compact
\ion{H}{2} region powered by a B0.5-type star; e.g.,
\citealp{hughes1982});

ii) a distribution of YSOs that follow the curvature of the structure
and is abundant in the boundary between the Ridge and the cluster
\citep{rivera2011}; and

iii) an overall much younger population compared to all the other YSO
groups in the rest of the field \citep{rivera2011}.  Indeed, the AFGL
333 Ridge contains $\sim70$\% of the Class 0/I population in the AFGL
333 field but only $\sim5$\% of the Class II population (this census
excludes the population in the East Loop whose environmental
conditions are more consistent with the western fields than the HDL;
Rivera-Ingraham et al. 2013, in preparation).

W3 SE and the AFGL 333 Ridge are both forming stars \citep{rivera2011}
and both appear to have the potential to form high-mass stars. They
also have a possible `trigger', i.e., a high-mass star, in their local
neighborhood that could aid in the process. However, they have not
reached the column densities, masses, and degree of stellar activity
(internal and external to the clumps) characterizing W3 East and W3
West.

\section{Formation of Clusters with High-Mass Stars and
  `Trapezium-like' Systems by `Convergent Constructive
  Feedback'}\label{sec:msf-conclusion}

A very interesting point noted in previous studies of both W3 East and
W3 West (e.g., \citealp{tieftrunk1997}; \citealp{bik2011}) is a
progressive \textit{decrease} in age of the stellar population from
the outskirts toward the peak of the density structure, from the
farthest diffuse \ion{H}{2} regions, to more centrally located evolved
compact, and HC\ion{H}{2} regions.
This is a critical clue.
The physical properties inferred from \herschel\ for W3 East and W3
West are unique in the entire W3 GMC.   In order to interpret them in the
context of their (also unique) geometry, stellar population
characteristics, and star formation history, we propose a scenario
for formation of a massive clump suitable for hosting a cluster of
high-mass stars as well as for formation of the individual high-mass
stellar members.  We now describe the key features of what we call
\textit{`convergent constructive feedback'}.

\subsection{Key Features}\label{key}

\subsubsection{Feedback}\label{feedback}

Low and high-mass star formation induced (`triggered') by external OB
stars is a well-studied phenomenon supported by extensive theoretical and observational studies (e.g.,
\citealp{white1999}; \citealp{tothill2002}; \citealp{minier2009}).
Local stellar feedback could result in much faster and more efficient star
formation than in the quiescent mode at the interaction boundaries
between the `triggering' stars and a dense environment (e.g.,
\citealp{elmegreen1977}).

We might have observed this effect between W4 and the HDL, near AFGL 333
(e.g., \citealp{rivera2011}; Rivera-Ingraham et al. 2013, in
preparation).  Enhanced star formation in triggered regions has also
been reported in previous studies (e.g., \citealp{thompson2012};
\citealp{koenig2012}).  

However, the progress of triggering is predicted to be dependent on the environmental conditions.  A high-mass star in a dense environment (e.g., a shell or ridge) can only induce further compression only at smaller
(sub-parsec) scales; the disruptive effects of newly formed high-mass
stars are not efficient in relatively dense regions \citep{dale2011}
and are therefore of more limited range.  Nevertheless, this further
compression could propagate the triggering process and the formation
of new high-mass stars on the relevant small scales, starting at the
boundaries and progressing toward the more dense (inner) regions of
the compressed structure.  This process would explain, for instance,
the presence of molecular cores and condensations at the perimeter of
the ionized regions (e.g., \citealp{tieftrunk1995};
\citealp{tieftrunk1998b}) in W3 Main (W3 East).

While (low/intermediate-mass) star formation likely has taken place in
the dense region prior to the present triggering, accounting for the
large cluster of low mass stars in the region \citep{feigelson2008},
the rate and efficiency will likely be enhanced at later stages. The
effects of sub-parsec triggering acting within the high column density
structure being formed, together with the amount of mass and limited
range of the triggering, could sustain lasting periods of star
formation in the most central regions, therefore emphasizing the
differential age effect.

\subsubsection{Constructive}\label{constructive}

The `positive' effects of stellar feedback by high-mass stars have
been studied extensively in the Galaxy, including other HOBYS fields (e.g., \citealp{zavagno2010}; Minier et al. 2013, A\&A, in press).
Here we argue that high-mass stars (`triggers') can collectively
influence not only the creation of new high-mass stars, but also the
new, massive structures hosting this new population of high-mass
stars.
Indeed, it is a requirement of our scenario that in the dense
environment, the progressive formation of high-mass stars will in
addition result in the creation of even higher column densities by
feedback, rather than simply disruption and dispersal by their
mechanical and radiative output.

This is in contrast to alternative models of high-mass cluster
formation, like that presented in \citet{peters2010}. In the scenario presented by these authors a central star forms first, followed by the formation of secondary stars in the accretion flow (i.e., star formation propagates outward).

\subsubsection{Convergent}\label{convergent}

To achieve the constructive behavior, we suggest that the
\textit{configuration} of the high-mass triggers is a key factor in
the formation of the most compact and massive systems (i.e.,
Trapezium-like systems), like those in W3 Main.  When acting on a
dense region with enough mass, different populations of high-mass
stars with the right `confining' configuration can lead to a
`convergent' process, creating/enhancing a central massive structure,
moving/trapping the material, and ensuring the availability of mass
for accretion during the early otherwise disruptive stages of
high-mass star formation.  The central column density would continue
to grow as new high-mass stars form in a sequential process by
sub-parsec triggering at the `boundaries' of the high column density
region, where triggering is most effective.

A particularly favorable case could arise where a massive and dense
structure is formed \textit{between} separate high-mass star populations,
due to their combined effect in compressing and \textit{confining} the
material.
W3 West exemplifies this possibility.  This prominent quiescent column
density peak has infrared sources at its periphery, with a clear
anticorrelation between molecular and ionized gas (e.g.,
\citealp{tieftrunk1998}).  A similar confining arrangement is also
observed for W3 East, and furthermore, higher resolution studies have
shown an even smaller central core containing IRS5, within our
\herschel\ clump, surrounded by the other \ion{H}{2} regions (e.g.,
\citealp{tieftrunk1995}), suggesting that W3 East contains a smaller
scale (and already active version) of what is occurring in W3
West. In addition, \citet{tieftrunk1998} suggested that the lack of
ammonia associated with W3 East was due to NH$_3$ being `thinned out'
or destroyed by ongoing activity \textit{without} dispersal, which
also supports the `confinement' aspect of the process.  In contrast,
an ammonia peak is still present in W3 West. The combined evidence
supports a similar process for both clumps, but with W3 West at an
earlier stage.

The concept of convergence is critical for high-mass star/cluster
formation as an active process aiding in the continuing supply of
material, beyond that required to form a low-mass star.  Convergence
might arise for a variety of different reasons.  Other than our
scenario above, evidence for different geometries is presented in
recent \herschel\ HOBYS studies for the importance of convergence of
flows \citep{hennemann2012} and of junctions of filaments
\citep{schneider2012}.
In addition to those scenarios aiming to describe the origin of the
parsec-scale progenitors of massive clusters, there are various models
describing the origin of the individual cluster members.  These invoke
processes acting at sub-parsec scales, among them small-scale
convergence of flows \citep{csengeri2011} in addition to turbulent
cores \citep{mckee2003}.

The compression of convergent feedback might also be able to explain
the `pinched' morphology of observed magnetic fields (e.g.,
\citealp{roberts1993}; \citealp{greaves1994};
\citealp{tieftrunk1995}), with an enhancement of the component of the
field parallel to the compressed structure (e.g.,
\citealp{peretto2012}).

\subsubsection{The combination in W3 Main}\label{combination}

A key consequence of the `convergent constructive feedback' process is
that stars would form progressively closer to the central regions,
each generation `aiding' in the formation of new high-mass stars, and
resulting in a systematic \textit{decrease} in age toward the
innermost regions of the clump.  Returning to our motivation, this
behavior is indeed observed in the high-mass stellar population in W3
Main (e.g., \citealp{tieftrunk1997}), as well as in the enhanced
concentration of young Class 0 stars at the boundaries of the
\ion{H}{2} regions within the W3 East clump \citep{ojha2009}
suggestive of induced star formation.  Thus the new scenario can
account for the unique stellar distribution and characteristics of W3
East and W3 West (spatial, age, and mass distributions, and multiplicity) in conjunction with the \herschel-based properties.
A sequential process of high-mass star formation in W3 Main was
already suggested in previous studies (e.g., \citealp{tieftrunk1997};
\citealp{feigelson2008}: option 4 in their analysis).

\subsection{Implications}\label{discussion}

When trying to address even the most basic processes of the high-mass
star formation process, current theoretical models face several
challenges such as:

i) the low core accretion rate $m^{*}$ (and therefore the long
formation times) due to initially low temperatures in the case of the standard protostellar collapse model (e.g., $m^{*}
\propto T^{3/2}$; for isothermal core collapse in the case of spherical collapse; \citealp{shu1977});

ii) the suppression of accretion due to radiation pressure and
ionization by the forming high-mass star \citep{zinnecker2007};

iii) formation in clustered environments; and

iv) primordial mass segregation with anomalous age distributions (e.g.,
young central massive systems surrounded by a cluster of older
low-mass stars).

The `convergent constructive feedback' process might provide a useful
framework for addressing some of these outstanding problems in
high-mass star formation.

First, given the age spreads in W3 observed by \citet{bik2011}, a
progressive formation of the central column density clump and cluster
members must have occurred over a $2-3$\,Myr period.  Therefore, there
is no need to form such structures fast enough to prevent major
internal fragmentation or to invoke long `starless'
lifetimes (e.g., several free-fall times; \citealp{mckee2003}), which allow for the material to be gathered before the star formation process is initiated (e.g., \citealp{zinnecker2007}).  The `older' halo cluster of low-mass stars surrounding the
high-mass star population in W3 Main (e.g., \citealp{megeath1996};
\citealp{feigelson2008}) that formed throughout the region might
already have initiated or enhanced the process of compression in the
center, as well as contributed to the formation of the first
population of high-mass stars.

Second, the simultaneous, small-scale (sub-parsec) triggering by
high-mass stars could provide more turbulent, as well as warmer
environments. The latter in particular could inhibit fragmentation by increasing
the minimum Jean's mass, leading to the formation of new massive cores
(e.g., \citealp{zinnecker1993}; \citealp{mckee2003};
\citealp{peters2010}) and an increase in characteristic stellar mass
toward the more central regions, as observed for the IRS5 clump for
the high and low-mass population \citep{megeath1996,ojha2009}.  The
combination of high efficiency of triggering and higher temperatures
could then be key to the formation and tell-tale characteristics of
rich clusters of high-mass stars.  The final morphology of the cluster
would resemble that of a more evolved cluster after mass segregation.
Indeed, from their timescale analysis of W3 Main, \citet{ojha2009}
suggested that the apparent mass segregation must not be dynamical in
origin.

Third, when a high-mass star forms close to the central (and most
dense) regions of the clump, the limited range of the stellar
influence \citep{dale2011} and the efficiency of triggering could then
lead to the most compact and richest systems, by forming new
overdensities, inducing the collapse of preexisting ones, or by direct
interaction between the effects of the embedded high-mass stars such
as outflows and shocks \citep{phillips1988}. In W3 East, this could
account for the high star formation efficiency and multiplicity
observed in the innermost regions of the clump, local to the IRS5
system (\citealp{megeath2005}; \citealp{rodon2008}). The sub-parsec
convergence of flows scenario from \citet{csengeri2011} would also
benefit from the confined environment created by the convergent
constructive feedback and/or converging flows, minimizing the disruptive effects.

Fourth, in a scenario with multiple and simultaneous triggering by
various high-mass stars, as in W3 Main, the resulting central
structure could reach rare, high column densities suitable for the
formation of central Trapezium-like systems.

Fifth, the continuing confinement and influence by the high-mass stars
at the outer boundaries of clumps could aid the accretion required to
produce very massive protostars in the central region, sustaining the
feeding process by counteracting or minimizing mass loss due to the
stellar outflows, winds, and radiative energy of the newly formed (and
more embedded) high-mass stars. How the actual feeding (accretion)
mechanism at in-clump scales proceeds (e.g., in filaments) is,
however, beyond the scope (and resolution capabilities) of this work.

In addition to the examples in the W3 GMC, `convergent constructive
feedback' might also be responsible for the observed morphology and
high-mass star formation in other regions, like the S255-S257 complex
\citep{minier2007}.

\subsection{W3 (OH) and Other High Column Density Regions}\label{confine}

W3 (OH) has an asymmetric cluster spread \textit{along} the tangent to
the direction of IC 1795, evidence of direct triggering by IC 1795
itself (e.g., \citealp{oey2005}; \citealp{feigelson2008}). This
influence of the IC 1795 is supported by the `string' of stellar
clusters extending north of W3 (OH) in the same tangential
orientation.  There is, however, also evidence of the influence of W4
on the structure containing all of these systems, as observed in the
various nebulosities and stars extending from W4 toward the W3 (OH)
`ridge'/shell \citep{tieftrunk1998}, and in the dynamics of this
structure and the associated stellar population (e.g.,
\citealp{thronson1985}).

The location of W3 (OH) at the interaction point between the effects
of two high-mass populations is therefore more reminiscent of large
(parsec) scale convergence of flows, albeit in this case directly
powered by high-mass stars rather than large-scale cloud
turbulence. Indeed, very large-scale feedback (e.g., over tens of
parsecs) might by itself be already capable of enhancing and inducing
the collapse of pre-existing structures (e.g., \citealp{peretto2012}; Minier et al. 2012 submitted).

Whether `convergence of flows' or `convergent constructive feedback' are responsible,
it appears based on the derived \herschel\ properties that similar
clumps are produced.  However, contrary to W3 Main, W3 (OH) has not
benefited from the combined effect of favorable environmental
conditions, the efficiency of sub-parsec triggering by (boundary)
high-mass stars, and the associated positive effects.  This lack of
local high-mass stars might explain why, for similar clump mass, the
W3 (OH) system is overall less active, and/or why there are two
massive clumps with significant high-mass star activity in W3 Main,
compared to just one in the `shell/ridge' containing W3 (OH).

On the other hand, W3 SE and the AFGL 333 Ridge have only relatively
weak and/or \textit{localized} (one-sided and of scale $<1$\,pc)
high-mass star feedback, driven by isolated late O/early B stars that
likely originated from larger-scale stellar feedback from IC 1795 and
W4, respectively.  Although small-scale (sub-parsec) triggering might
have led to the formation of these high column density structures, the
collective effects of compression, confinement, and stellar feedback
have not (yet) been sufficient to form high-mass stars within them.

\section{Conclusion: A Requisite of High-Mass Star Formation Theories}\label{requisites}

This work is the first analysis of the W3 GMC using the \herschel\
data obtained as part of the Guaranteed Time Key Program HOBYS.

The data were reduced, corrected for contributions from
background/foreground material, and used to produce and analyze the
column density and dust temperature maps. The software
\textit{getsources} was used to extract a preliminary catalog of
compact sources from the column density maps and the \herschel\ maps
convolved to the $500$\,\micron\ resolution.  In particular, the
multiband datasets were used to carry out a detailed study of the
intrinsic (SED) properties of the clumps currently hosting on-going
high-mass star formation, suitable for comparison with the results and
suggested high-mass star formation models from other HOBYS fields.

Our results indicate that the $\sim0.4$\,pc clumps hosting on-going high-mass star
formation are unique in the W3 GMC in terms of luminosity, temperature, mass,
column density, and stellar content, with W3 East, W3 West, and W3 (OH) being the most massive and
the highest column density regions of the entire GMC.

W3 East and  W3 West, in addition to W3 (OH), are the only clumps
currently forming clusters of high-mass stars in this field.
Therefore, we have used the properties \textit{exclusive} to these
clumps to develop a scenario linking their `extreme' \herschel-based
properties with the well-known characteristics of their high-mass
stellar content, which are also unique within the W3 GMC.

While numerical models and simulations are required to test the
feasibility of different scenarios, we conclude that the observational
evidence in W3 points toward a very basic requisite for high-mass star
formation: an active and continuing assembly of material for the
formation of the most massive cores/clumps.  In particular, and in
agreement with the conclusions from other HOBYS studies, `convergence'
appears to be a common feature for achieving this goal, distinguishing
this from low mass star formation.  The movement of material would
guarantee a build-up of mass and ensure maintenance of the feeding
(accretion) process of the massive protostars during their earliest
stages despite their disruptive power.  A dynamical formation of
massive precursors has also been suggested in previous studies based
on a statistical analysis of core populations \citep{motte2007}.

A major difference between the convergence of flows scenario and the
scenario proposed here is that we use stellar feedback, the efficiency
of triggering in star formation, and the associated displacement,
compression, and confinement of material, to ensure the availability
of mass during cluster formation.  In this process of `convergent
constructive feedback,' the sub-parsec influence from different
populations of high-mass stars can lead to the creation/enhancement of
a massive clump in the most central regions, a suitable precursor of
the most massive Trapezium-like systems.  In addition there are
sequential star formation events initiated at the boundaries of the
high-column density structure.  This scenario could explain not only
the non-dynamical cases of mass segregation, but also those clusters
with stellar age decreasing toward the innermost regions of the
cluster.

The power, configuration, scale of the triggering process, and the
environmental conditions are all likely important in determining the
final richness, geometry, timing (e.g., inner young regions versus an
outer older population), and the IMF of the formed cluster. Overall,
the low probability of satisfying all the prerequisites and conditions
for the formation of the most massive clusters could underlie why
there are only relatively small numbers of systems like those in
Orion, W3 IRS5, and the massive clumps in DR 21.

\acknowledgements 
AR-I acknowledges support from an Ontario Graduate Scholarship and a
Connaught Fellowship at the University of Toronto. 
This research was supported in part by the Natural Sciences and
Engineering Research Council of Canada and the Canadian Space Agency
(CSA). The authors also thank the anonymous referee for his/her useful comments.
SPIRE has been developed by a consortium of institutes led by Cardiff
Univ. (UK) and including: Univ. Lethbridge (Canada); NAOC (China);
CEA, LAM (France); IFSI, Univ. Padua (Italy); IAC (Spain); Stockholm
Observatory (Sweden); Imperial College London, RAL, UCL-MSSL, UKATC,
Univ. Sussex (UK); and Caltech, JPL, NHSC, Univ. Colorado (USA). This
development has been supported by national funding agencies: CSA
(Canada); NAOC (China); CEA, CNES, CNRS (France); ASI (Italy); MCINN
(Spain); SNSB (Sweden); STFC, UKSA (UK); and NASA (USA).  
PACS has been developed by a consortium of institutes led by MPE
(Germany) and including UVIE (Austria); KU Leuven, CSL, IMEC
(Belgium); CEA, LAM (France); MPIA (Germany); INAF-IFSI/OAA/OAP/OAT,
LENS, SISSA (Italy); IAC (Spain). This development has been supported
by the funding agencies BMVIT (Austria), ESA-PRODEX (Belgium),
CEA/CNES (France), DLR (Germany), ASI/INAF (Italy), and CICYT/MCYT
(Spain).
D.P. is funded through the Operational Program "Education and Lifelong Learning" and is co-financed by the European Union (European Social Fund) and Greek national funds. 
K.L.J.R. is funded by an ASI fellowship under contract number I/005/11/0.

\clearpage

\appendix


\section{\textit{Herschel} Images of W3}\label{5images}

\begin{figure}[ht]
\centering
\subfigure[PACS $70$\,\micron]{%
\label{fig:log70.eps}
\includegraphics[scale=0.37,angle=270]{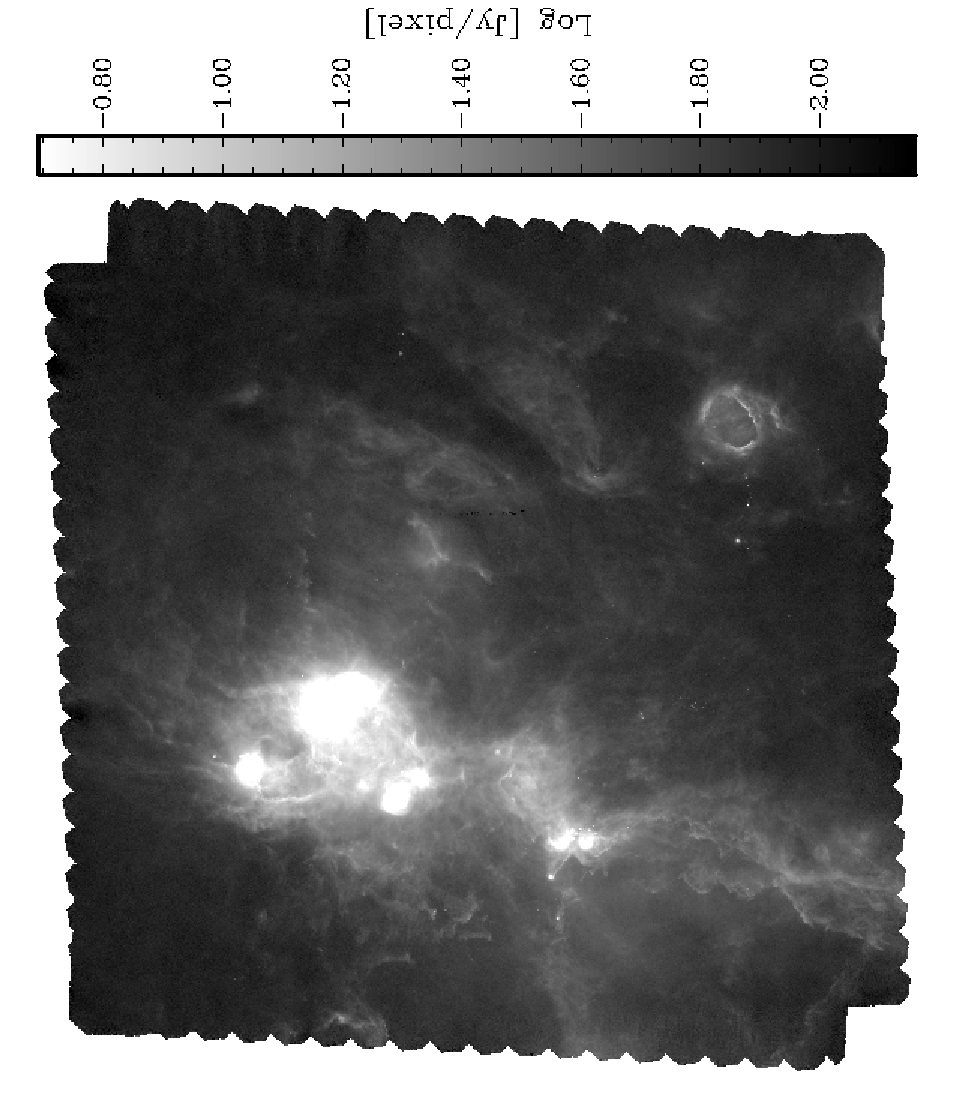}
}%
\subfigure[PACS $160$\,\micron]{%
\label{fig:log160.eps}
\includegraphics[scale=0.37,angle=270]{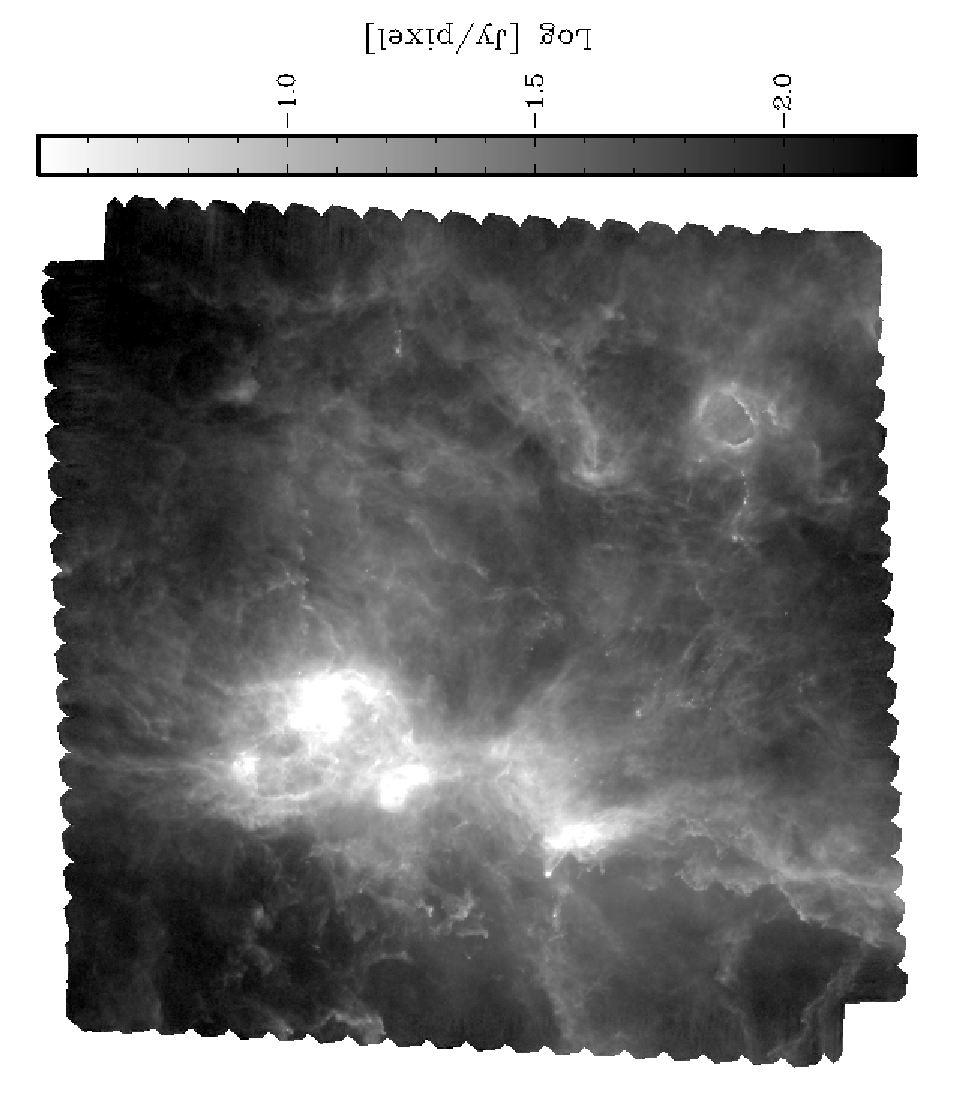}
}\\%
\subfigure[SPIRE $250$\,\micron]{%
\label{fig:log250.eps}
\includegraphics[scale=0.37,angle=270]{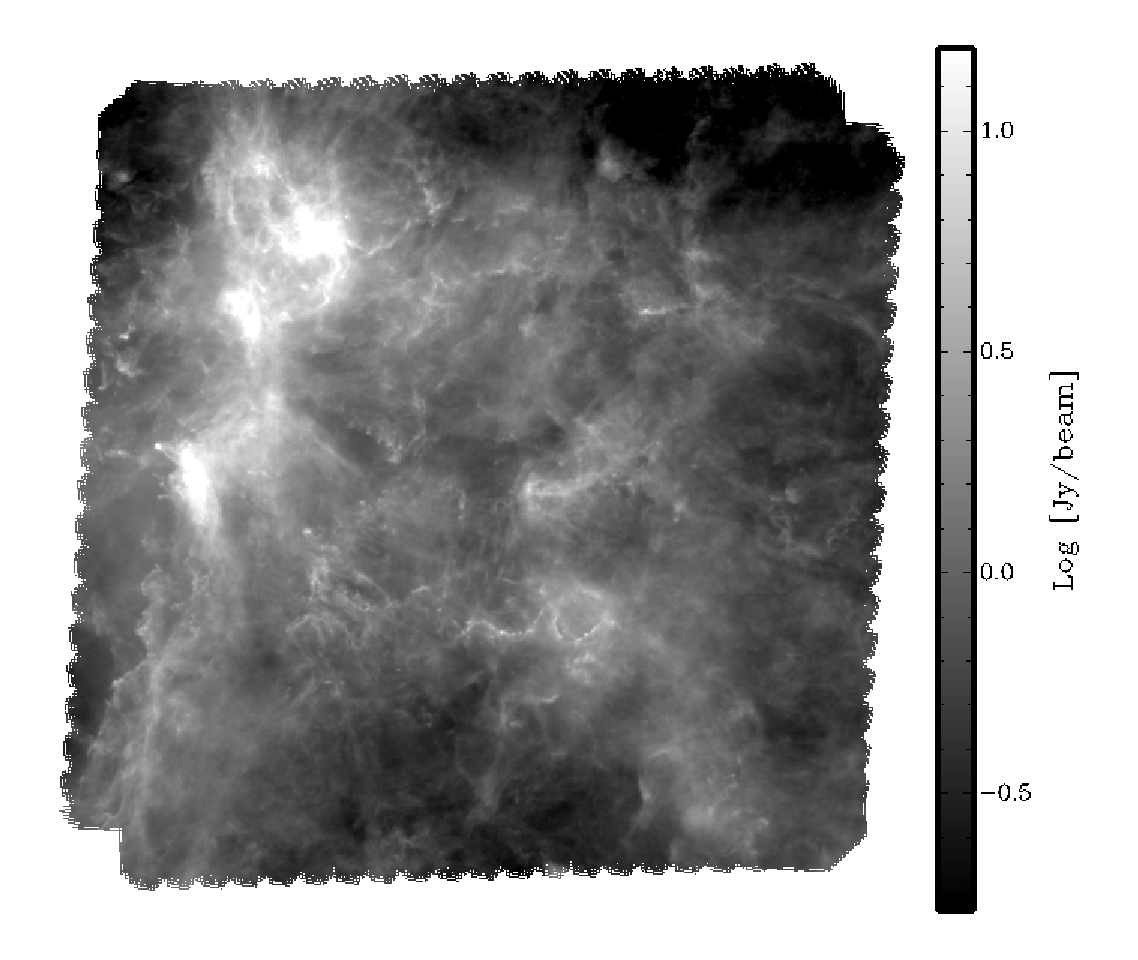}
}%
\subfigure[SPIRE $350$\,\micron]{%
\label{fig:log350.eps}
\includegraphics[scale=0.37,angle=270]{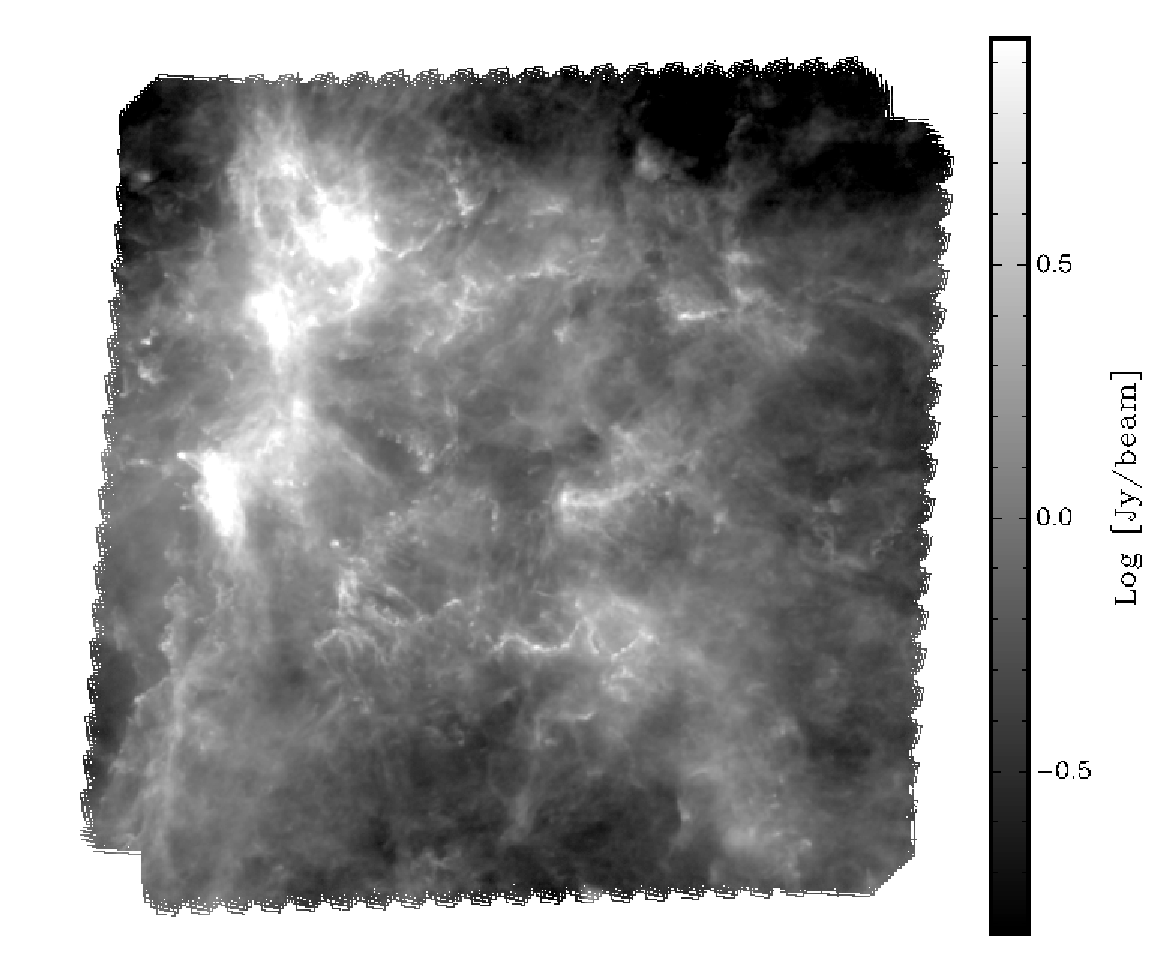}
}\\%
\subfigure[SPIRE $500$\,\micron]{%
\label{fig:log500.eps}
\includegraphics[scale=0.37,angle=270]{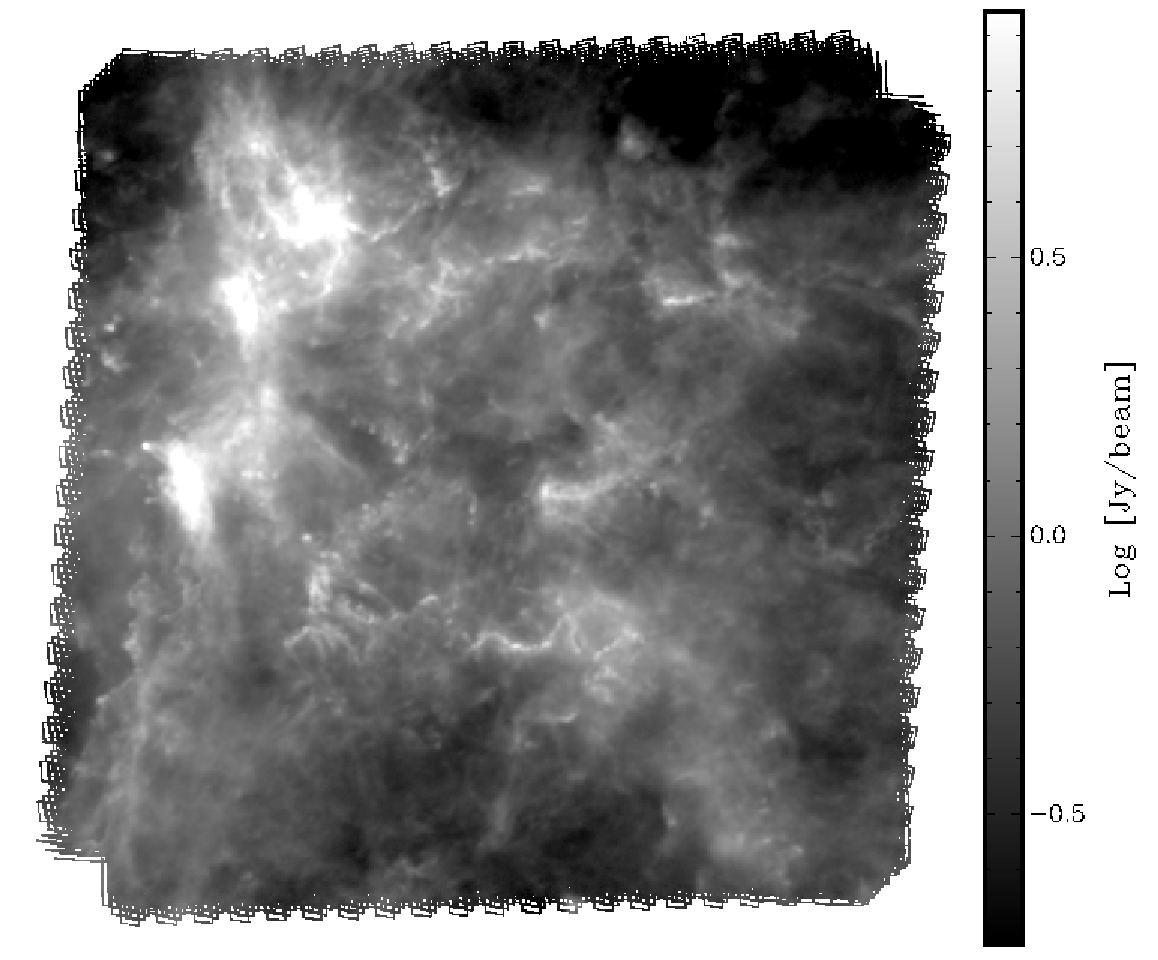}
}%
%
\label{fig:monos}
\end{figure}

\clearpage


\section{Contributions to the \textit{Herschel} Intensities by Dust Associated with
  Foreground/Background Gas}\label{foreground}

While emission in the \herschel\ intensity maps is dominated by the W3
GMC, these maps are still affected by contributions from background
and (mainly) foreground material not local to the W3 GMC. To correct
for this emission, as done in, e.g., \citet{planck2011Td}, we need to
estimate the contribution $I_{\nu,i}$ at each \herschel\ band for two
contributing components: dust in foreground/background gas that is
traced by H~I emission (atomic; subscript a) and CO emission
(molecular, subscript m).  The optically-thin approximation to
Equation~\ref{column} is sufficient except for the rare high column
density peaks where the foreground/background emission is
inconsequential, and so $I_{\nu,i} = \sigma_{\nu,i} B_\nu(T_i)
N_{{\mathrm H},i}$.

We used the \ion{H}{1} and CO molecular spectral data cubes from the
CGPS to estimate the foreground and background contribution.  From
visual inspection of the CO cube, the velocity range of the W3 GMC is
about $-29$ to $-60$\,km\,s$^{-1}$.
The foreground/background gas column was therefore estimated by
integrating the respective line emission over velocity,
\emph{excluding} this range, giving us $W({\rm H\, I})_a$ and $W({\rm
CO})_m$. 

For the atomic component, we determined $N_{{\mathrm H},a}$ by the
standard conversion of $W({\rm H\, I})_a$ (e.g.,
\citealp{lequeux2005}), ignoring self absorption which would be an
uncertain and generally small correction.  The few regions that are
highly absorbed against the continuum of very bright \ion{H}{2} regions were
replaced in the column density map by an estimate of the emission
using a centered annulus 3\arcmin\ wide; in any case these tend to be
high total column density regions where the foreground/background
correction is not important.

For the molecular component, we made a preliminary estimate
$N_{{\mathrm H},m} = 2 X_{\rm CO} W({\rm CO})_m$ adopting $X_{\rm CO}$
from \citet{strong1988}.  In the final analysis, it is actually the
product $\sigma_{\nu,m} X_{\rm CO}$ that is needed, and we calibrate
this below.  

The CGPS resolution is only $\sim1$\,\arcmin\ compared to the
36\arcsec\ required, and these maps have to be interpolated onto
the common 9.0\,\arcsec\ grid of the preliminary \herschel-based
$N_{\mathrm{H_2}}$ and $T$ maps.

\subsection{Parameters for the Foreground/Background Atomic Component}

To determine a dust temperature $T_a$ representative of the atomic
gas, we selected regions with $W({\mathrm{CO}}) <
0.5$\,K\,km\,s$^{-1}$, low column density in the preliminary column density map (column density below the mean value of the map), the ratio of atomic to molecular ISM contribution greater than one, and with the foreground/background column density greater than the estimated column density in the W3 GMC by at least a factor of five.  An example of such a location is the upper
left region of the column density map. Pixels in this region of the temperature map are characterized by $T_a \sim16$\,K, consistent with the lower limit of the temperature range for the
atomic material as estimated by \citet{planck2011Dd,planck2011Td}.

For atomic gas, $\sigma_{\nu,a}(1200)$ at our fiducial frequency
$\nu_0$ ranges somewhat about a mean $\sim 1 \times 10^{-25}$
\,cm$^2$\,H$^{-1}$ and $T \sim18$\,K with $\beta=1.8$
\citep{planck2011Dd,planck2011Td}.  Due to the relatively uniform
value of $N_{{\mathrm H},a}$ over the field, we were unable to
calibrate $\sigma_{\nu,a} (1200)$ ourselves toward this field and so
we adopted this value and $\beta=1.8$.

With these parameters, we estimated $I_{\nu,a}$ across the entire
field, and subtracted them from the observed \herschel\ intensities.
As a consistency check, we observed that in the regions dominated by
the atomic foreground/background, there is no oversubtraction.

\subsection{Parameters for the Foreground/Background Molecular Component}

To determine a dust temperature $T_m$ representative of the molecular
gas, we again selected regions of overall low column density and with the foreground/background column density greater than the estimated column density in the W3 GMC by at least a factor of five. Now, however, we required the ratio of atomic to molecular ISM contribution to be less than one and $W({\mathrm{CO}})_m > 3$\,K\,km\,s$^{-1}$.  An example of such a
location is the region below the West Loop.  At the positions of
these regions we estimated $T_m \sim 15$\,K from the preliminary temperature map.

By correlating $I_\nu - I_{\nu,a}$ (smoothed to 1\arcmin) with
$W({\mathrm{CO}})_m$, we found the slope $\sigma_{\nu,m}
X_{\mathrm{CO}} B_\nu(T_m)$ frequency by frequency.  This provides a
consistency check on the adopted $T_m$ and $\beta = 1.8$.  The product
$\sigma_{\nu,m} (1200) X_{\mathrm{CO}}$ was found to be in good agreement
for all bands $\ge160$\,\micron\ (those used to create the column
density maps), with a mean value
$[3.6 \pm 0.3] \times 10^{-5}$\,H$^{-1}$ (K\,km\,s$^{-1}$)$^{-1}$.  We
adopted this empirical calibration to estimate $I_{\nu,m}$ across the
entire field.  Again we checked for oversubtraction.

If $X_{\mathrm{CO}} = [2.3 \pm 0.3] \times 10^{20}$\, molecules cm$^{-2}$
(K\,km\,s$^{-1}$)$^{-1}$ from \citet{strong1988}, then $\sigma_{\nu,m}
= [1.6 \pm 0.2] \times10^{-25}$\,cm$^2$\,H$^{-1}$.
\citet{planck2011Td} estimated $\sigma_{\nu,m} (1200)$ to be $2.3
\times 10^{-25}$\,cm$^2$\,H$^{-1}$ for the molecular phase in the
Taurus molecular cloud, within the broad range found in other
environments by \citet{martin2012}.  If this were the value of the
opacity, then
$X_{\mathrm{CO}} = [1.6 \pm 0.2] \times 10^{20}$\,molecules cm$^{-2}$
(K\,km\,s$^{-1}$)$^{-1}$.  Recent results from \emph{Fermi}
\citep{abdo2010} in the second quadrant find
$X_{\mathrm{CO}} = [1.9 \pm 0.2] \times 10^{20}$\,molecules cm$^{-2}$
(K\,km\,s$^{-1}$)$^{-1}$ in the Perseus arm,
$[1.59 \pm 0.17] \times 10^{20}$\,molecules cm$^{-2}$
(K\,km\,s$^{-1}$)$^{-1}$ in the Local arm, and
$[0.87 \pm 0.05] \times 10^{20}$\,molecules cm$^{-2}$
(K\,km\,s$^{-1}$)$^{-1}$ in more local Gould Belt clouds.  For Local
arm clouds in the third quadrant,
$X_{\mathrm{CO}} = [2.08 \pm 0.11] \times 10^{20}$\,molecules
cm$^{-2}$ (K\,km\,s$^{-1}$)$^{-1}$ \citep{ack2011}.

\subsection{Results}

We subtracted the above estimates of $I_{\nu,a}$ and $I_{\nu,m}$ from
the \herschel\ intensity maps.  We then repeated the pixel by pixel
SED fitting process using Equation~\ref{column} to find the column
densities and dust temperatures characteristic of the W3 GMC itself.
For the GMC material, we again adopted the optical depth to column
density conversion described in Section~\ref{sec:maps}.

On average over the pixels in the map, the foreground/background
correction decreases the observed column densities by $\sim30\pm15$\%,
with a greater proportional effect on the lower column density regions
and little effect on the high column density peaks and ridges.

We note that some `artifacts' might inevitably appear in areas in
which the foreground/background dominates: after the subtraction the
residual intensity is quite small, whereas the errors remain, so that
it is difficult to obtain a reliable fit.  The main region suffering
from this effect is located south of the West Loop, where the
relatively prominent foreground/background CO component was used to
calibrate the opacity.  Here, the intensity correction has produced a
prominent feature in Figure \ref{fig:temp-ism} near
2$^{\mathrm{h}}$\,22$^{\mathrm{m}}$\,50$^{\mathrm{s}}$,
$+61^{\circ}$\,0\arcmin\,40\arcsec, which also produces the spike in
$T$ at low column densities seen in the panel for KR 140 in
Figure~\ref{fig:coltemp_histo_ism2}.  Such regions could be masked for
any analysis sensitive to details at low column densities,
particularly because these are most prone to other systematic errors,
including uncertainties in the offsets added to the \herschel\ maps
derived from \emph{Planck} and \IRAS, uncertainties in the map-making
techniques, use of the same $\sigma_{\nu,i}$ and $T_i$ for all of the
dust in the foreground and background gas components, and optical
depth effects in the gas tracers across the field.


\bibliographystyle{apj}

\end{document}